\documentclass[journal=mamobx,manuscript=article]{achemso}

\makeatletter
\def\@dotsep{4.5}
\makeatother

\usepackage[version=3]{mhchem}
\usepackage{amsmath}
\usepackage{amsfonts}
\usepackage{amssymb}
\usepackage{bm}
\usepackage{latexsym}
\usepackage{float}
\usepackage{graphicx}
\usepackage{color}
\usepackage{multirow}
\usepackage{sidecap}

\author{Ali Makke}
\email{ali.makke@epf.fr}
\affiliation{Universit\'e de Lyon- Univ. Lyon I - LPMCN - UMR CNRS 5586- F69622 Villeurbanne, France}
\alsoaffiliation{Universit\'e de Lyon - INSA Lyon - MATEIS - UMR CNRS 5510 - F69621 Villeurbanne, France}
\alsoaffiliation{Ecole Polytechnique F\'eminine, Institut Charles Delaunay - LASMIS UMR CNRS 6279- F10004 Troyes, France}

\author{Olivier Lame}
\email{olivier.lame@insa-lyon.fr}
\affiliation{Universit\'e de Lyon - INSA Lyon - MATEIS - UMR CNRS 5510 - F69621 Villeurbanne, France}

\author{Michel Perez}
\email{Michel.perez@insa-lyon.fr}
\affiliation{Universit\'e de Lyon - INSA Lyon - MATEIS - UMR CNRS 5510 - F69621 Villeurbanne, France}

\author{Jean-Louis Barrat}
\email{jean-louis.barrat@ujf-grenoble.fr}
\affiliation{Universit\'e Joseph Fourier, Grenoble - LIPhy - UMR CNRS 5588- F38402 Grenoble, France}

\title{Nano-scale buckling in lamellar block polymers:\\ a molecular dynamics simulation approach}

\begin{document}


\begin{abstract}

Oriented block copolymers exhibit a buckling instability when submitted to a tensile test perpendicular to the lamellae direction.
In this paper we study this behavior using a coarse grained  molecular dynamics simulation approach.
Coarse grained models of lamellar copolymers with alternate glassy rubbery layers are 
 generated using the Radical Like Polymerization method, and their mechanical response is studied. 
For large enough systems, uniaxial tensile tests  perpendicular to the direction of the lamellae reveal the occurrence of the  buckling instability at low strain.
The results that emerge from molecular simulation are compared to an  elastic theory  of the buckling instability introduced by Read and coworkers. 
At high strain rates, significant differences are observed between  elastic theory and simulation results for the  buckling strain and the buckling wavelength.
We explain this difference by  the strain rate dependence of the mechanical response.
A simple model that takes into account the influence of the strain rate in the mechanical response is presented to rationalize the results at low and moderate strain rates. 
At very high  strain rates,  cavitation takes place in the rubbery phase of the sample and  limits the validity of the approach.  

\end{abstract}

%

\section{Introduction}
\label{sec:intro}

Block copolymers such as (Styrene-Butadiene-Styrene SBS or Styrene-Isoprene-Styrene SIS) have attracted much interest in the past few decades for their use as  thermoplastic elastomers. The microstructure of such materials results from the mixture of two different homo-polymers. Interesting combination of properties (similar to the case of nano-composites) at ambient temperature can be obtained when one of the constitutive homo-polymer is hard (glassy or crystalline) while the other one is soft (rubbery).
As both constituents are linked together by chemical cross-links, the resulting material combines the mechanical properties of each phase: therefore ductility of the rubbery phase  is coupled to the toughness of the glassy phase.
 
The mechanical response of such composite materials is far from being understood, especially at the molecular scale. Depending on the amount of each component, the thermodynamics equilibrium between phases leads to various morphologies (\emph{e.g.} spherical, cylindrical and lamellar~\cite{Bates99}). The lamellar morphology is particularly interesting as a model system because the global behavior is not dominated by one of the component. Moreover, the one dimensional aspect of the lamellar morphology is similar to the morphology of semi-crystalline polymers at small scale. 

Such nano-structured materials exhibit similar mechanical behavior ~\cite{Michler04,Krumova06,Koo05,Hermel03} through complex deformation mechanisms when they are stretched perpendicular to the lamellae direction: the hard lamellae buckle, forming a ``chevron-like'' morphology. This phenomenon has been observed in semi-crystalline polymers by several authors ~\cite{Burnay77,Blochl85} and more recently by Bartczak and Mohanraj ~\cite{Bartczak052,Mohanraj08}. Buckling of hard lamellae is of prior importance regarding the mechanical properties of nanostructured polymers: it induces a rapid collapse of the hard phase network which can be an initiator of the macroscopic yield, decreasing thus the mechanical properties of the material.

\begin{figure*}
\includegraphics[width=0.8\textwidth]{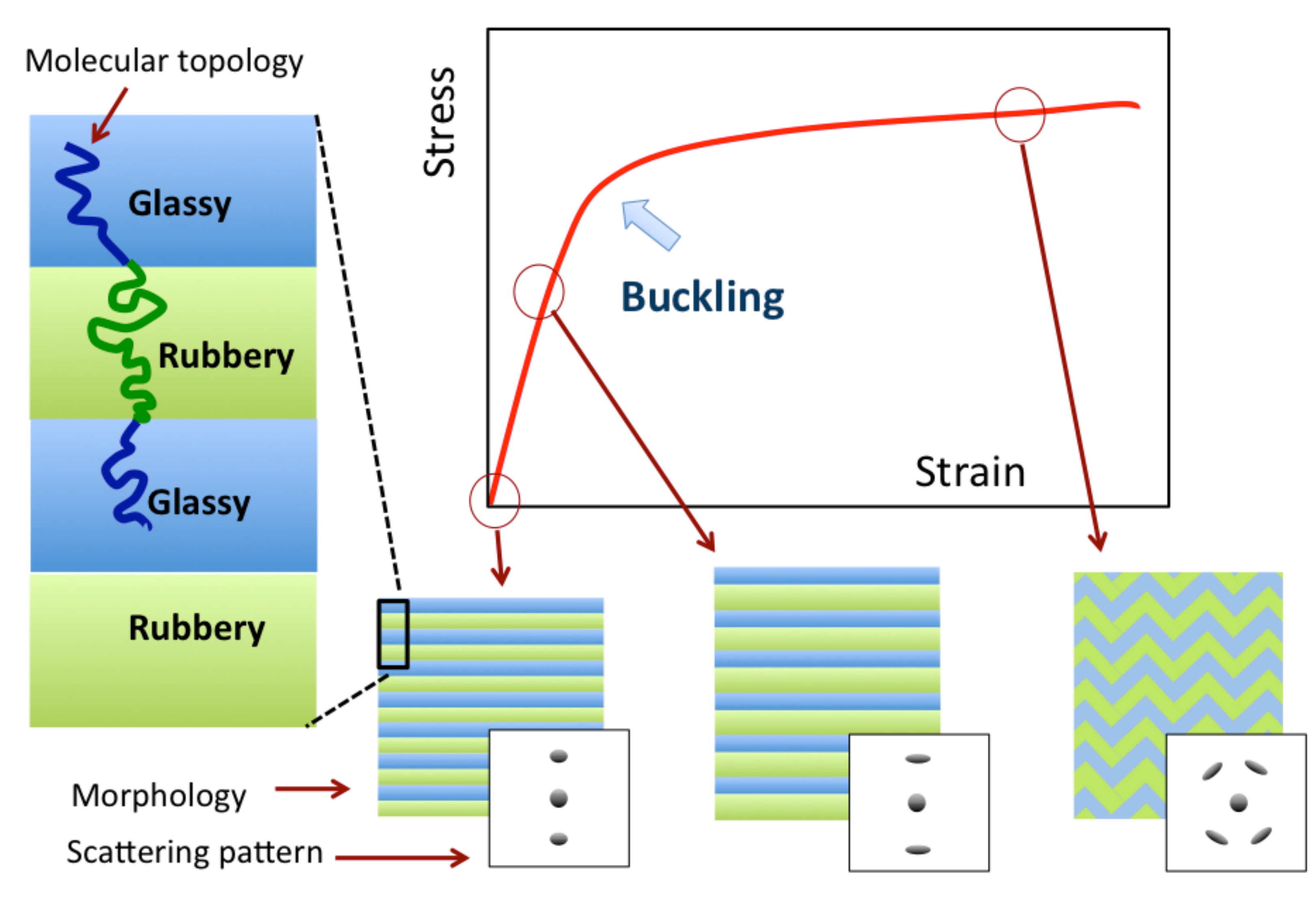}
\caption{Schematic description: Evolution of a triblock morphology from lamellar to chevron-like morphology under  tensile test conditions. This transformation separates two regimes: (i) the elastic regime, (ii) the buckling regime. A schematic description of the phases topology is shown at left. The corresponding SAXS pattern are shown in the inset. }
\label{SBSbehavior.fig}
\end{figure*}

Several experimental works~\cite{Cohen00,Mori03,Mahanthappa08,Phatak06,Honeker06,Diamant88} were focused on the formation of ``chevron''  in block-copolymers using \emph{in situ} Small Angle X-ray Scattering (SAXS). The results reveal a progressive modification of the SAXS pattern at yield, from two spots to four symmetric spots, which is the signature of the ``chevron'' morphology (see ~\ref{SBSbehavior.fig}).
 
Buckling was also observed by a direct examination of the microstructure of highly deformed SBS samples using atomic force or electron microscopy~\cite{Adhikari04,Cohen_1_01}. The origin of buckling is interpreted from the contrast in the elastic properties of the glassy and rubbery layers. As the rubbery layer accommodates most of the imposed deformation, it tends to contract in the transverse direction. The strong coupling between phases leads to a compressive stress in the transverse direction of the glassy layer. As a consequence, the glassy lamellae buckle to form a ``chevron'' morphology. 

Theses qualitative explanations are well established. However, the attempts to model the phenomenon theoretically are scarce,  and essentially at the level of a continuum description. In a  pioneering work, Read \emph{et al}~\cite{Read99} proposed an original energetic approach to describe the competition between buckling and affine deformation, and proved with Finite Element Methods (FEM) that this instability can occur even in the pure elastic regime. Unfortunately,  this purely elastic approach ignores  visco-elastic or rate dependent effects , and does not account for the competition with other failure modes such as cavity nucleation.

In this paper, we study the initiation of the buckling instability and the mechanical response of a block copolymer model using coarse grained molecular dynamics simulations. This numerical tool is indeed particularly adapted to study this problem as it intrinsically contains elasticity, viscosity, and all associated dynamical effects. It also accounts for defects, and for the different coupling strengths that can exist between phases, depending on the density of  tie molecules~\cite{Makke12}. Moreover, coarse grained MD permits the study of buckling well above the yield stress in the plastic regime, where other instabilities such as cavitation can occur~\cite{Makke09}. The  approach gives  access to the local measurement of many variables (\emph{e.g} local stress, local density) while monitoring the global mechanical behavior. This allows  one to investigate the relationship between the mechanical response and the change in the local microstructure and morphology. 

This paper is a companion paper of a previously published report~\cite{Makke11_2}, which showed the feasibility of such and approach and gave a preliminary account of the results. Section II presents the model and the methods that are used in this study. Section III discusses the relationship between the microstructure and the associated stress-strain curve. Section IV recalls the formalism introduced by Read\emph{et al}~\cite{Read99}. The following sections detail the effect of sample size (section V) and strain rate (section VI). A simple kinetic model predicting the competition between buckling modes is finally presented in section VII and compared with MD results.

\section{Method and model}
\label{sec:SimTech}

\subsection{The molecular dynamics model}

Molecular dynamics (MD) simulations were carried out for a well established coarse-grained model~\cite{Kremer90}, in which the polymer is treated as a chain of $N$ beads, which we refer to as monomers, of mass $m=1$ connected by a spring to form a linear chain. The beads interact with a classical Lennard-Jones (LJ) interaction :

\begin{equation}
\mathrm{U_{LJ}^{\alpha\beta}(r)}=\left\{ \begin{array}{ll}
4\epsilon_{\alpha\beta}\left[\left(\frac{\sigma_{\alpha\beta}}{r}\right)^{12}-\left(\frac{\sigma_{\alpha\beta}}{r}\right)^{6}\right] & \mbox{, \ensuremath{r\le r_{c}}}\\
0 & \mbox{, \ensuremath{r> r_{c}}}\\
\end{array}\right.
\label{LJpot}
\end{equation}

\noindent where the cutoff distance $r_{c}=2.5\sigma$. $\alpha$ and $\beta$ represent the chemical species (\emph{i.e.}  $A$,  $B$) In addition to \eqref{LJpot}, adjacent monomers
along the chains are coupled through the well known anharmonic Finite Extensible Nonlinear Elastic potential (FENE):

\begin{equation}
\mathrm{U_{FENE}(r)}=
-0.5kR_{0}^{2}\ln{\left[1-\left(\frac{r}{R_{0}}\right)^{2}\right]}  \mbox{, \ensuremath{r\le R_{0}}}
\label{FENEpot}
\end{equation}

\noindent The parameters are identical to those given in Kremer~\emph{et al}~\cite{Kremer90}, namely $k=30\epsilon/\sigma^{2}$ and $R_{0}=1.5\sigma$, chosen so that unphysical bond crossings and chain breaking are avoided. All quantities will be expressed in terms of length $\sigma$, energy $\epsilon$, pressure $\epsilon/\sigma^{3}$ and time $\tau=\sqrt{m\sigma^{2}/\epsilon}$.

Newton's equations of motion are integrated with the velocity Verlet method and a time step $\Delta t=0.006\tau$. Periodic simulation cells containing $340\times n$ chains ($n$ is the replication number - see below) of $N=200$ beads were used with a Nos\'e-Hoover barostat, \emph{i.e.} in the NPT ensemble. An anisotropic barostat with $P_{x}=P_{y}=P_{z}=0$ is used in the equilibration, leading to a tetragonal simulation box before running the tensile
test.

\begin{figure*}
\includegraphics[width=\textwidth]{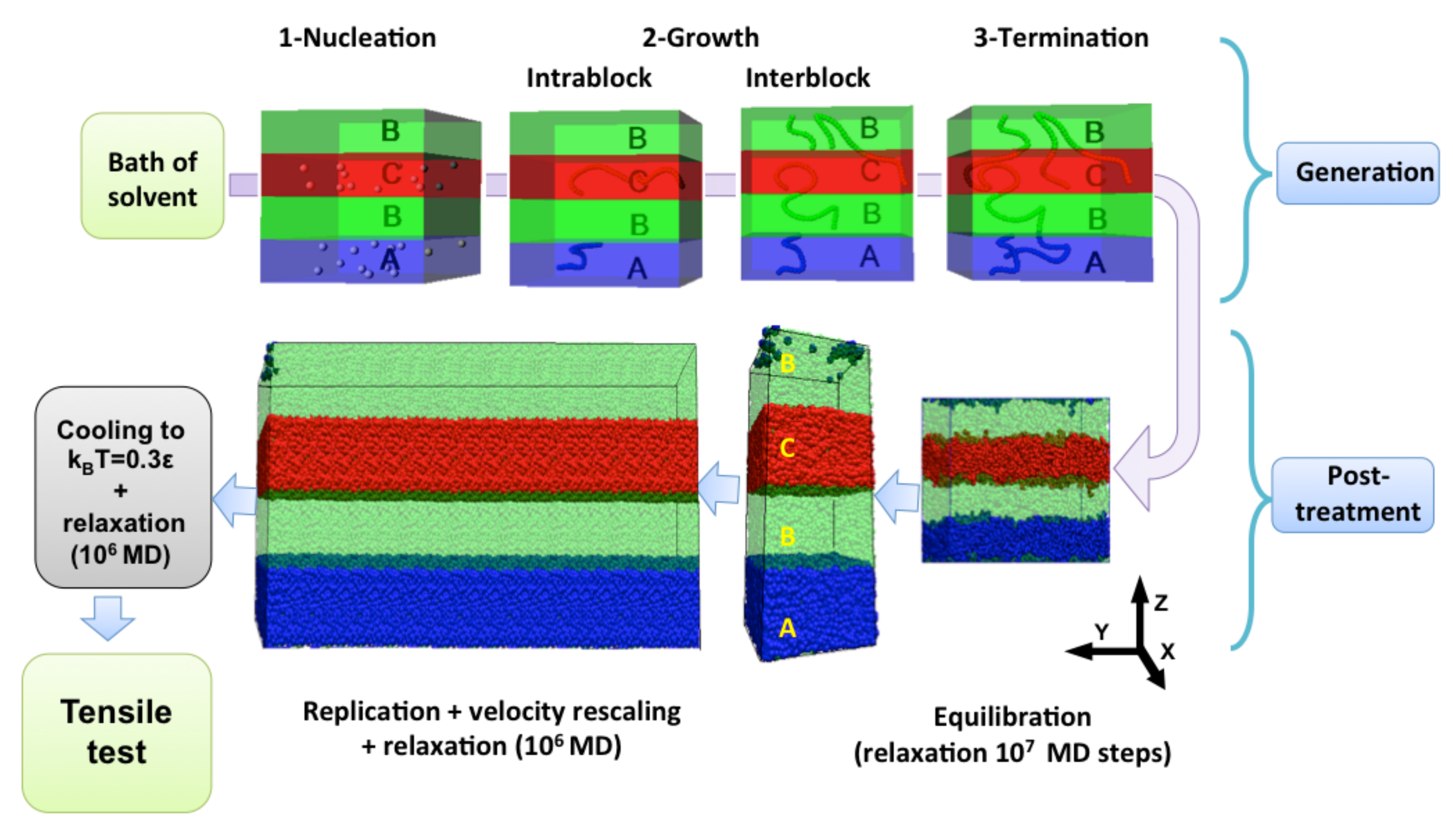}
\caption{Schematic description of the generation and the post treatment process. Starting from a bath of monomers (''solvent''), the Radical Like Polymerization routine will transform the single disconnected beads to entangled chains. The polymerization occurs in 4 blocs $A$, $B$, $C$ and $B$. The resulting chains are shared between three successive blocks $A$, $B$ and $C$. LJ interaction energies are chosen so that a perfect segregation of the blocks is ensured at low temperature.}
\label{RLPgeneration.fig}
\end{figure*}

\subsection{Polymer sample generation}
Our samples have been generated using the ``Radical-Like Polymerization'' (RLP)  method~\cite{Perez08}. The polymerization takes place in a LJ liquid bath (solvent)  for which $\epsilon_{\alpha\beta}=1~\epsilon$ and $\sigma_{\alpha\beta}=1~\sigma$ for all LJ interactions. A set of 340 monomers are chosen as ``radicals''. The growth of chains is then performed by a sequence of bead-addition (growth) and subsequent relaxation. In the growth stage, each radical captures one of its free nearest neighbors and  a covalent bond is created between them. The newly  bonded  monomer becomes itself a radical. After that, the system is relaxed during 500 MD steps at $k_BT=1~\epsilon$. Then, a new growth step is performed, until chains reach the desired size. 

This concept was adapted to generate triblock copolymer samples with four distinct blocks $ABCB$ and four interfaces parallel to  the $(xy)$ plane. The generation starts simultaneously in blocks $A$ and $C$. When the chains reach a length of $N/4$, the radical is dragged  towards the nearest interface to complete the growth of chain in neighboring blocks $B$. After the chain reaches a length of $3N/4$, the radical, that started growth in block $A$ moves to achieve the polymerization in block $C$ and \emph{vice-versa}. The generation is stopped when all chains attain the request length $N$. The generation method is detailed in references~\cite{Perez08, Makke12}. 

\subsection{Equilibration and replication}
After the  sample has been generated, the remaining solvent is removed from the simulation box. The LJ interaction energies are then  adjusted to drive the segregation and associated block thickness. In this study, the LJ energies are chosen such that $A$ and $C$ layers are glassy, while $B$ layer is rubbery: $\epsilon_{AA}=\epsilon_{CC}=1~\epsilon$, $\epsilon_{BB}=0.3~\epsilon$ and $\epsilon_{AB}=\epsilon_{BC}=0.4~\epsilon$. The system is then relaxed $10^7$~MD steps in NPT ensemble at $k_BT=1~\epsilon$. All the pressure components are maintained at zero ($P_x=P_y=P_z=0$) using an anisotropic barostat, allowing box changes in the three dimensions independently. The evolution of the box lengths during the relaxation steps has been measured and it has been found that the box dimensions reach a steady state after $10^7$ MD steps, indicating that mechanical equilibrium  is reached.

In order to study the effect of sample size, larger samples were built by replicating several times the basic periodic sample in the $y$ direction. The replication was performed at $k_BT=1 \epsilon$ where the two phases are rubbery. To avoid unphysical internal periodicity, bead velocities are rescaled and an additional $10^{6}$~MD steps are performed. After this relaxation stage, each sample is  cooled down to a  temperature of $k_BT=0.3\epsilon$ during $7\times10^{5}$~MD steps and relaxed again $10^{6}$~MD steps. Four different box sizes (in the $y$ direction) have been investigated: 200, 400, 500 and 800~$\sigma$. 

Finally, the glass transition temperatures of each phase were determined from volume curves  at constant pressure, and it was found that $T_g^A= 0.43$ and $T_g^B= 0.20$. In the following the system will be  studied at a temperature intermediate between $T_g^A$ and $T_g^B$, so that the $B$ phase has the properties of a rubber and the A phase those of a glass.   \ref{RLPgeneration.fig} describes schematically the sequence operations for both generation and post treatment stages of the sample.

\subsection{Tensile test:}

To deform our samples, uniaxial homogenous tensile test conditions were employed~\cite{Makke09}. The samples were subjected to a sequence of deformation-relaxation steps, composed of: (i) a rescaling of the simulation box in the tensile direction ($Z$ in our case, so that the true strain is  $\epsilon_{zz}(t)=\ln(L_z(t)/{L_z(0)}$); and, (ii) one MD step in the NPT ensemble at $k_BT=0.3~\epsilon$ and $P_x=P_y=0$ (Nos\'e-Hover anisotropic barostat is employed to control the pressure only in $x$ and $y$ directions independently). The tensile velocity $\dot{L}_z$  was chosen so that the initial strain rate is $\dot{\epsilon}_{yy}(0)=7.3\times10^{-5}\tau^{-1}$ (high strain rate tests) and $\dot{\epsilon}_{yy}(0)=1.4\times10^{-5}\tau^{-1}$ (low strain rate tests).

\section{ stress/strain curve}
\label{sec:StressStrain}
\begin{figure*}
\setlength\unitlength{1cm}
\begin{center}
\includegraphics[width=0.48\textwidth]{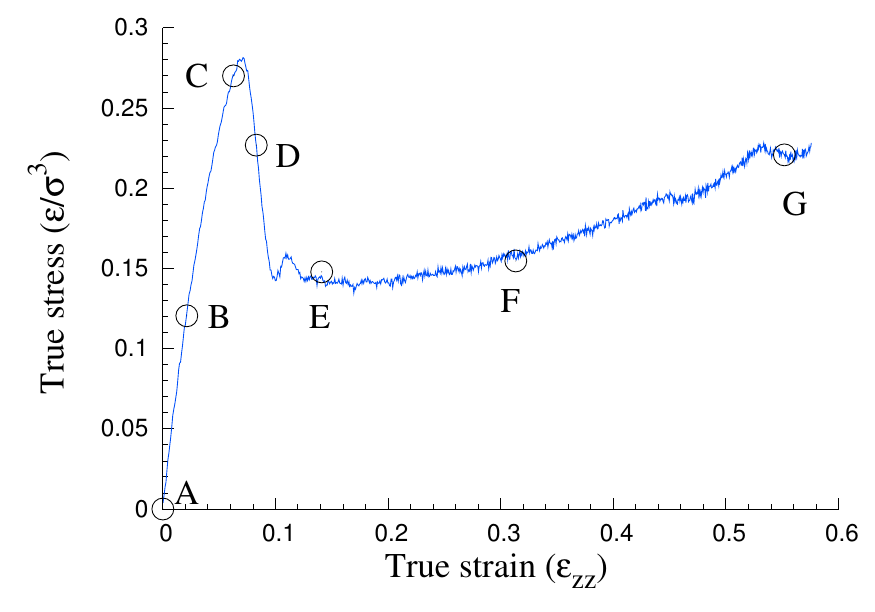}
\includegraphics[width=0.48\textwidth]{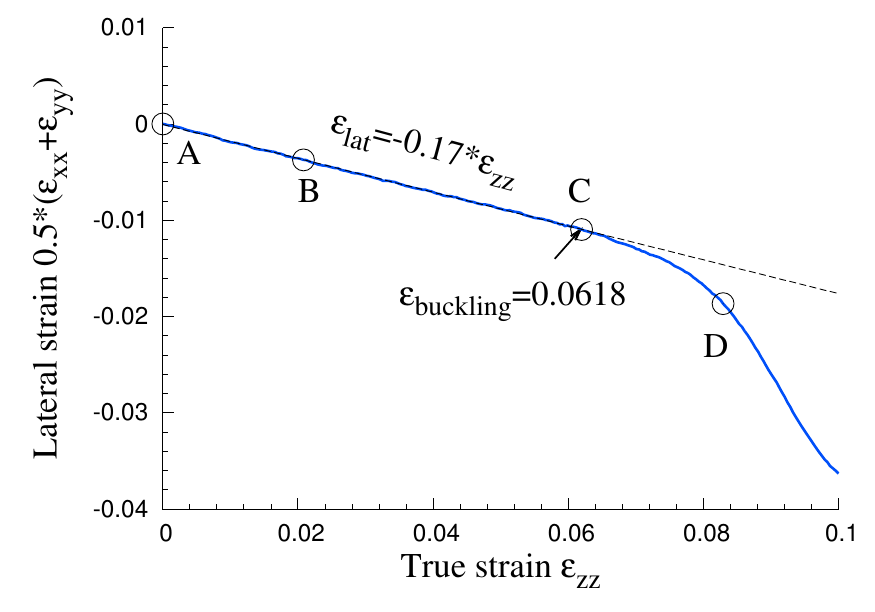}
\end{center}
\caption{ Left-panel: Stress strain curve of a sample submitted to a uniaxial tensile test in the direction normal to  the layers ($Z$ direction). The black circles correspond to the density maps  shown in figure \ref{fig:DensityMap}. Right panel: Lateral strain in the  same sample with respect to the tensile strain. The first linear part of the curve fits the Poisson ratio of the sample in the elastic regime. The curve deviates from this linear behavior after  buckling has occurred.}
\label{fig:BehaviorCurveH}
\end{figure*}

\begin{figure}
\setlength\unitlength{1cm}
\begin{center}
\includegraphics[width=0.5\textwidth]{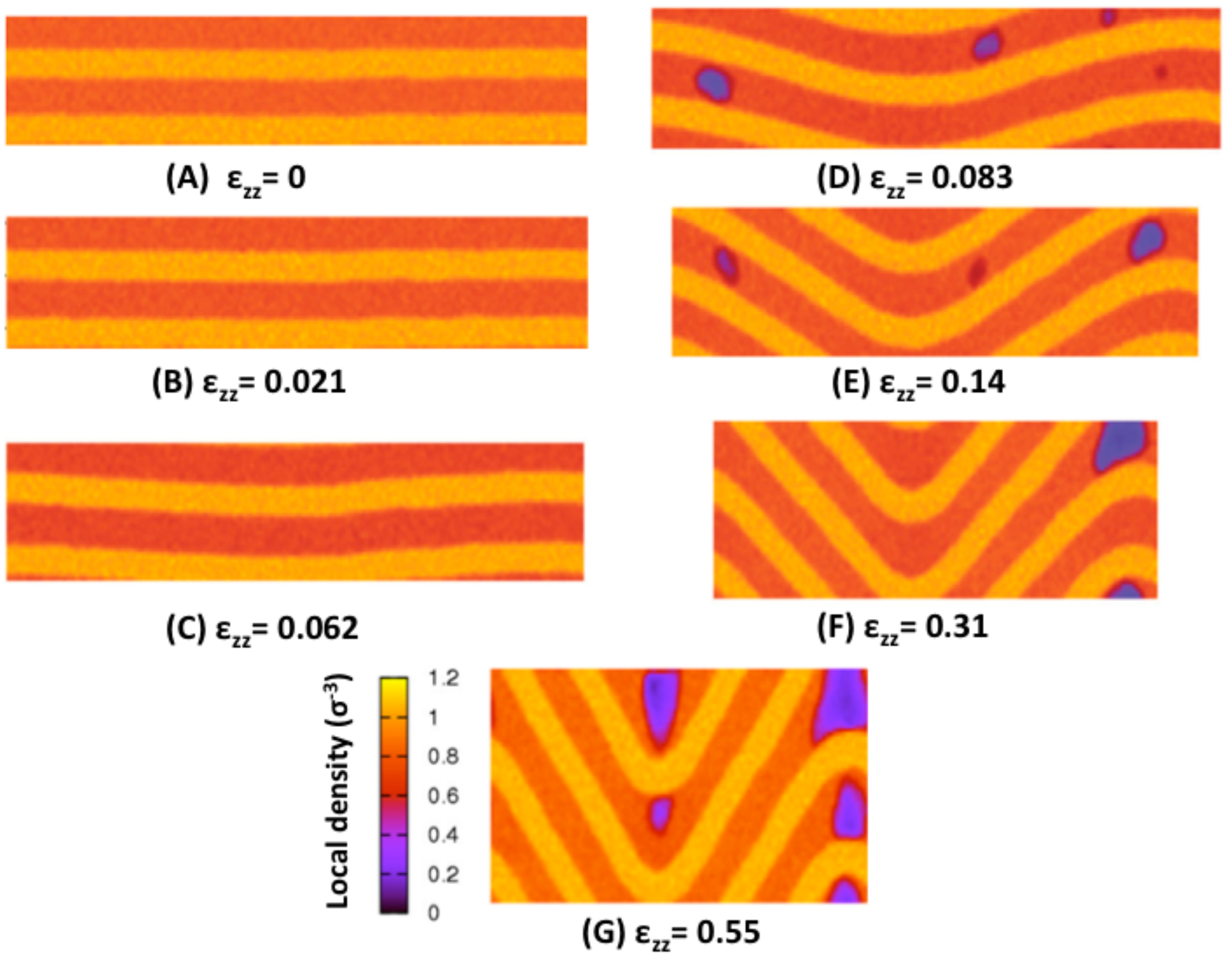}
\end{center}
\caption{ Local density cartography of the tested sample at several strains. The color contrast illustrates the alternance of  glassy and rubbery lamellae with high and low density, respectively.  As the deformation progresses, nucleation of  cavities occurs randomly in the rubbery phase. Cavities that are located in the tilted part of the chevron disappear, as the deformation in these regions becomes a simple shear deformation; however, only cavities that are located in the hinges, where the deformation is effectively triaxial,  persist to a high strain. }
\label{fig:DensityMap}
\end{figure}

In order to study the correlation between the mechanical response and the change of the sample morphology, we start by presenting the results of  uniaxial tensile tes on ``large'' ($L_Y\simeq 400\sigma$) samples. The tensile strain was applied in the $Z$ direction (perpendicular to the lamellae) at a constant velocity $V_z=\dot{L_z}$. In this first set of results, the strain rate is $\dot{\epsilon_{zz}} = 7.3\times10^{-5} \tau^{-1}$ and the initial size of the simulation box is $(32.4\times74.2\times393.6)~\sigma^3$. The resulting stress-strain curve is plotted in figure  \ref{fig:BehaviorCurveH}.  During the tensile test, sample configurations were stored  at different strains. Figure \ref{fig:DensityMap} shows the local density map of such configurations, where the glassy and the rubbery phases can be distinguished by the high and low density lamellae, respectively.

The stress-strain curve in  figure \ref{fig:BehaviorCurveH} exhibits different regimes.
In the first elastic regime the stress growth linearly. 
This response results from the serial coupling of the lamellae and the imposed strain is mainly accommodated by the deformation of the soft phase (rubbery).
Beyond the elastic regime a progressive softening is observed. 
This softening is correlated to a progressive change in the morphology of the  sample, as 
 the aligned lamellae start to buckle, leading to an undulated pattern.
 Buckling is signaled by a  drastic change of  the apparent Poisson ratio, illustrated in Figure \ref{fig:BehaviorCurveH}-right , which displays
  the lateral strain as a function of  the imposed normal  strain.
 The first linear part of the curve corresponds to the linear  elastic regime, and the slope gives the apparent Poisson ratio of the sample.
A sudden  decrease of the Poisson ratio appears when the buckling starts.
After the onset of  buckling, cavities nucleate in the rubbery phase, leading thus to a strong drop in the  stress. 
Figure \ref{fig:DensityMap} shows the nucleation and the evolution of the cavities (low density spots in the rubbery phase).
With increasing strain the amplitude of the buckling undulation develops.  The cavities that are first randomly nucleated in the rubbery phase,become progressively  localized at the hinges of the pattern.
Indeed, cavities nucleated in the tilted part of the buckle tend to heal, as the stress in these regions becomes a simple shear. On the contrary, the stress  at the hinges is essentially triaxial, and favors cavitation. 

 The micro-mechanical  origin of buckling instability has been elucidated in reference \cite{Makke11_2}. 
Due to the serial coupling between the glassy and rubbery phases, the tensile strain will be mainly localized in the rubbery lamellae.
 The tensile strain in the normal direction of the layers will be converted locally to a contraction in the transverse direction because of the Poisson effect in the rubber.
This lateral contraction will be transmitted to the glassy phase via the interface, then a compressive stress acts on the transverse direction of the glassy lamellae.
Under these conditions,  and for a sufficiently large system, a buckling instability takes place to relax the lateral compressive stress.

\section{Theoretical modeling of buckling instability}
\label{sec:ElasticTheo}

The development  of a buckling instability in a layered material with alternative hard and soft blocks can be studied using elasticity theory. The volume average elastic energy density of a homogeneously strained  sample is, as usual,  given by:
\begin{equation}
\langle e\rangle = \frac{1}{2} C_{ijkl}\langle\epsilon_{ij}\epsilon_{kl}\rangle 
\end{equation}
where $\epsilon_{ij}$ are the components of the macroscopic strain, and $C_{ijkl}$ the elastic constants. 
For the sake of simplicity, only a 2D case is presented here: the $x$ direction is parallel to the lamellae, and $z$ the direction of traction, perpendicular tot the lamellae. Taking into account the symmetry of our system (transversely isotropic - see ~\ref{RLPgeneration.fig}), the volume averaged elastic energy density reduces to:
\begin{equation}
2 \langle e \rangle = C_{11}\langle\epsilon_{11}^2\rangle + C_{33}\langle\epsilon_{33}^2\rangle + 2C_{13}\langle\epsilon_{33}\epsilon_{11}\rangle +2C_{44}\langle\epsilon_{13}^2\rangle
\label{Energy.eq}
\end{equation}
where standard Voigt notations have been used. Here the energy density is expressed in a local frame that is aligned with the lamellar pattern, as will become clear below. The elastic constants are the effective values that describe the lamellar material as a whole, and depend from the characteristics of each phase.
A particularity of the  dibloc  material resides in the values of these elastic constants (see table~\ref{ElasticConstants.tab}). The material is remarkably soft when  submitted to shear in the $xz$ or $yz$ directions. Therefore, when submitted to a tensile stress in the $z$ direction, 
the material will  have a tendency to locally \emph{rotate} in order to align its soft directions at 45$^\circ$ from the  direction of traction.

\begin{table}
\caption{Elastic constants of the layered bock copolymer studied in this paper. Elastic constants have been determined with molecular statics on a stress-free sample by applying small perturbation in the box shape. For the sake of comparison, the Young modulus of a glassy polymer is of order $50~\epsilon/\sigma^3$}
\begin{center}
\begin{tabular}{c|c|c|c|c||c}
					&	$C_{11}$	&	$C_{33}$	&	$C_{23}=C_{13}$	&	$C_{44}=G$	&	$\nu$\\
\hline
[$\epsilon/\sigma^3$]		&	24		&	7.6		&	6.5				&	0.07		&	0.18
\end{tabular}
\end{center}
\label{ElasticConstants.tab}
\end{table}%

 The formalism of equation~\ref{Energy.eq}, which assumes a homogeneous strain,  is  not appropriate to  predict such a rotation, and in principle a full finite element calculation involving space varying elastic constants appropriate for the different phases would be required. The analysis proposed by 
Read \emph{et al}~\cite{Read99}  bypasses this difficulty, by describing  the local deformation of the sample $\epsilon_{ij}$ as a combination of a   ``shear+strain'' deformation  expressed in the local frame of the lamellae  (trough $\epsilon_{11}$, $\epsilon_{13}$ and  $\epsilon_{33}$ - transformation matrix $[\mathcal{S}]$) with a space dependent  rotation $\theta$ (transformation matrix $[\mathcal{R}]$, which describes the local tilt of the lamellar structure. The two transformation matrixes that describe these deformations  are:
\begin{equation}
[\mathcal{R}]= \left[\begin{array}{cc} \cos\theta&-\sin\theta\\ \sin\theta&\cos\theta\end{array}\right]\ \mathrm{and}\  [\mathcal{S}]=\left[\begin{array}{cc} 1+\epsilon_{11}&\epsilon_{13}\\ 0&1+\epsilon_{33}\end{array}\right] 
\label{RandS.eq}
\end{equation}

Under deformation, two  adjacent points $M_0(x_0,z_0)$ and $M(x_0+\delta x,z_0+\delta z)$  are transformed into $M_0'(x_0',z_0')$ and $M'(x'_0+\delta x',z'_0+\delta z')$, and  the vector $\mathbf {MM_0}$ experiences the combination of the rotation and deformation:
\begin{equation}
\mathbf{M'M_0'} = [\mathcal{R}] [\mathcal{S}] \mathbf{MM_0}
\label{OM_0'.eq}
\end{equation}
In terms of the coordinates, this reads: 
\begin{equation}
\left(\begin{array}{c} \delta x'\\ \delta z'\end{array}\right) = \left[\begin{array}{cc} \cos\theta&-\sin\theta\\ \sin\theta&\cos\theta\end{array}\right] \left[\begin{array}{cc} 1+\epsilon_{11}&\epsilon_{13}\\ 0&1+\epsilon_{33}\end{array}\right] \left(\begin{array}{c} \delta x\\ \delta z\end{array}\right) 
\label{Displacement.eq}
\end{equation}

If one now introduces the displacement field  $\mathbf{v}(x,z)$ such that $\mathbf{M_0'} = \mathbf{M_0} +\mathbf{v}(x_0,z_0)$, the relation between $\mathbf{M'M_0'}$ and $\mathbf {MM_0}$ is expressed in terms of the displacement gradients:



\begin{equation}
\left(\begin{array}{c} \delta x'\\ \delta z'\end{array}\right) = \left(\begin{array}{c} \delta x\\ \delta z\end{array}\right) +\left[\begin{array}{cc} \nabla_x v_x&\nabla_z v_x\\ \nabla_x v_z&\nabla_z v_z\end{array}\right]  \left(\begin{array}{c} \delta x\\ \delta z\end{array}\right) 
\label{Taylor.eq}
\end{equation}

Combining equations~\ref{Displacement.eq} and \ref{Taylor.eq} leads to a system of 4 equations that can be inverted to give $\theta$, $\epsilon_{11}$, $\epsilon_{13}$ and $\epsilon_{33}$ in terms of the  displacement gradient $ \nabla \mathbf{v}$:

\begin{equation}
\left\{\begin{array}{l}
\sin\theta=\nabla_xv_z/\Delta\\
\epsilon_{11}=\Delta-1\\
\epsilon_{33}=[(1+\nabla_zv_z)(1+\nabla_xv_x)-\nabla_zv_x \nabla_xv_z]/\Delta-1\\
\epsilon_{13}=[(\nabla_zv_x)(1+\nabla_xv_x)+(\nabla_xv_z)(1+\nabla_zv_z)]/\Delta
\end{array}\right.
\label{Epsilon.eq}
\end{equation}

where $\Delta=\sqrt{(\nabla_xv_z)^2+(1+\nabla_xv_x)^2}$. Note that if the sample is submitted to pure shear, such as only $\nabla_xv_z\neq0$, $\sin\theta=\epsilon_{13}$, \emph{i.e.} the shear is completely described  by the local rotation.

In order to describe an undulating pattern of the lamellar structure, the global displacement vector $\textbf{v}$ is decomposed in two contributions: the macroscopic deformation of the sample (strains $\epsilon_{xx}$, $\epsilon_{xz}$ and $\epsilon_{zz}$) and a small perturbation that imposes a displacement  along $z$ only: $\mathbf{u}=u_z(x)\mathbf{z}$:

\begin{equation}
\left\{\begin{array}{rcl}v_x&=&\epsilon_{xx}x\\
v_z&=&\epsilon_{zz}z+u_z\end{array}\right.
\label{Decomposition.eq}
\end{equation}
where $u_z(x)=U_0\sin(kx)$ is a sinusoidal small perturbation of wave vector $k$ that describes the local displacement due to buckling. 

Now, it is possible to calculate local deformations $\epsilon_{11}$, $\epsilon_{13}$ and $\epsilon_{33}$ of equation~\ref{Epsilon.eq} using the decomposition described in equation~\ref{Decomposition.eq}. Only second order terms of the small sinusoidal perturbation  $U_0$ are retained, leading to the following expressions for the  local deformations in the frame of the lamellae: 
\begin{equation}
\left\{\begin{array}{l}
\langle\epsilon_{11}^2\rangle=\epsilon_{xx}^2+\frac{U_0^2k^2}{2(1+\epsilon_{xx})^2}\epsilon_{xx}(1+\epsilon_{xx})\\
\langle\epsilon_{33}^2\rangle=\epsilon_{zz}^2-\frac{U_0^2k^2}{2(1+\epsilon_{xx})^2}\epsilon_{zz}(1+\epsilon_{zz})\\
\langle\epsilon_{11}\epsilon_{33}\rangle=\epsilon_{xx}\epsilon_{zz}+\frac{U_0^2k^2}{4(1+\epsilon_{xx})^2}(\epsilon_{zz}-\epsilon_{xx})\\
\langle\epsilon_{13}^2\rangle=\frac{U_0^2k^2}{2(1+\epsilon_{xx})^2}(1+\epsilon_{zz})^2
\end{array}\right.
\label{Epsilon.eq}
\end{equation}
It is worth noting that, for  a tensile test along the $z$ axis ($\epsilon_{xx}<0$ and $\epsilon_{zz}>0$), the rotation introduced previously induces a net decrease in the squared  local deformations, leading to an energy relaxation and demonstrating thus the possibility for an instability.

Inserting the identities above  into the energy density equation~\ref{Energy.eq} leads to the volume averaged elastic energy density:
\begin{equation}
2 \langle e \rangle = C_{11}\epsilon_{xx}^2 + 2C_{13}\epsilon_{xx}\epsilon_{zz}+C_{33}\epsilon_{zz}^2 
 +\frac{1}{2}U_0^2 k^2 f_1^{2D}
\label{Energybis.eq}
\end{equation}
where 
\begin{equation}
\begin{array}{lcl}
f^{2D} &= &[ G-\epsilon_{zz}( C_{33}- C_{13}-2G)- \epsilon_{zz}^2( C_{33}-G)\\
&& +\epsilon_{xx}( C_{11}(1+\epsilon_{xx})- C_{13})]/(1+\epsilon_{xx})^2
\end{array}
\label{f2D.eq}
\end{equation}
Due to the specific moduli of our composite system (see table~\ref{ElasticConstants.tab}), the term $f^{2D}$, slightly positive at zero strain, rapidly turns negative as strain increases (\emph{i. e.} after less than 1\% deformation in $z$). Thanks to this combination of transformations, the local shear in the $xz$ plane is completely handled by the rotation $\theta$, decreasing  the elastic energy of the system and causing thus the buckling instability. 

However, in equation~\ref{Energybis.eq}, there is no cost associated with the sinusoidal bending of the material, except the shearing of the soft phase, which is, as mentioned previously, very low. In order to account for the energy associated with the bending of the composite, and particularly the bending of the hard phase, a new term has to be added. It is assumed here, following the idea of Read~\emph{et al}, that only the hard layers contribute to the bending energy. A macroscopic expression is then introduced from beam mechanics, which gives the energy, that is necessary to bend a plate:
\begin{equation}
e_b=\frac{1}{2}K(\nabla_x\theta)^2
\label{bend_ene.eq}
\end{equation}
where $K$ is the bending modulus of the sample. Due to the serial coupling between phases the bending modulus will be dominated by the contribution of the  hard phase.  The bending modulus can  then be estimated from simple beam bending theory as $K=\phi_h^3 E_h d^2/[12(1-\nu_h^2)]$ where $\phi_h$ is the volume fraction of the hard phase, $E_h$ and $\nu_h$ are the Young's modulus and Poisson ratio of the hard phase and $d$ is the lamellar spacing.

The total energy density results from the addition of the bulk elastic energy and the bending energy, $\langle e_T\rangle= \langle e_b\rangle+\langle e\rangle$:
\begin{equation}
2\langle e_T\rangle=C_{11}\epsilon_{xx}^2 + 2C_{13}\epsilon_{xx}\epsilon_{zz}+C_{33}\epsilon_{zz}^2 +F(\epsilon_{zz},k)
\label{tot_energy.eq}
\end{equation}
with $F(\epsilon_{zz},\epsilon_{xx},k)=(U_0^2/2)(k^2 f^{2D} + K k^4)$. In the real 3D case, the same analysis can be performed, which leads to a function $f^{3D}$ instead of $f^{2D}$ where:

 \begin{equation}
f^{3D}=f^{2D} - \frac{C_{23} \epsilon_{yy}}{(1+\epsilon_{xx})^2}  
\label{f1_3D.eq}
\end{equation}
This leads to the final form of the function $F(\epsilon_{xx},\epsilon_{yy},\epsilon_{zz},k)$:
 \begin{equation}
 F(\epsilon_{xx},\epsilon_{yy},\epsilon_{zz},k)=\frac{U_0^2}{2}(k^2 f^{3D} + K k^4)
 \label{F.eq}
 \end{equation} 
 
The buckling instability occurs upon increasing strain when the function $F(\epsilon_{xx},\epsilon_{yy},\epsilon_{zz},k)$  in equation \ref{tot_energy.eq} becomes negative, meaning that the global gain in elastic energy overwhelms the bending energy penalty. 

Finally, before the buckling begins (\emph{i.e.}  in the elastic regime) $\epsilon_{xx}$ and $\epsilon_{yy}$ can be replaced by $\nu \epsilon_{zz}$ where $\nu$ is a global Poisson ratio. Under this assumption, $f^{3D}(\epsilon_{xx},\epsilon_{yy},\epsilon_{zz})$ becomes $f^{3D}(\epsilon_{zz})$ and $F(\epsilon_{xx},\epsilon_{yy},\epsilon_{zz},k)$ becomes $F(\epsilon_{zz},k)$.

\begin{figure}
\begin{center}
\includegraphics[width=\linewidth]{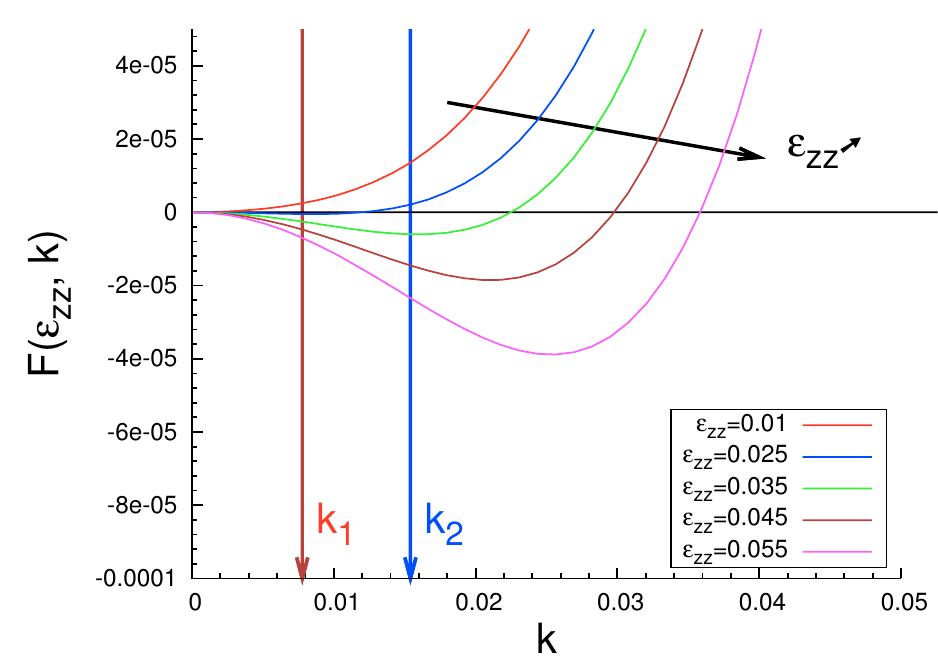}
\end{center}
\caption{Evolution of $F(\epsilon_{zz},k)$ (see equations~\ref{tot_energy.eq} and \ref{F.eq}) for different   levels of strain, the two arrows denote the wavevectors $k_1 = 2\pi/L_y$ and $k_2 = 4\pi/L_y$.
}
\label{fig:ElasticTheory}
\end{figure}

Figure \ref{fig:ElasticTheory} shows the evolution of  $F(\epsilon_{zz}, k)$ as a function of  the wave-vector $k$ for several tensile strains $\epsilon_{zz}$. At low strain $F(\epsilon_{zz}, k_n)$ is positive for any possible wave-vector $k$, therefore the buckling is impossible. As the tensile strain increases $F(\epsilon_{zz}, k_n)$ becomes negative for a wave-vector range, which indicates the possibility of buckling.

Note that the periodic boundary conditions imposed in the direction $y$ in our MD simulations limit the possible wavevectors to the discrete set:
\begin{equation}
k=k_n=n\frac{2\pi}{L_y}
\end{equation}
 where $n$ is the mode number; \emph{i.e.} $n=1$ is the fundamental mode where the wavelength of the perturbation is equal to the sample size. Figure \ref{fig:ElasticTheory} shows that as deformation increases, the first wavelength for which the sinusoidal perturbation might be stable is the fundamental one. As the deformation increases, the fundamental mode might not be the most energetically favorable, since larger wave-vectors lead to more negative values for $F(\epsilon_{zz}, k_n)$.

Finally, buckling occurs if:
 
 \begin{equation}
 f^{3D}(\epsilon_{zz}) + K \left(n\frac{2\pi}{L_y}\right)^2 < 0
\label{InstabilityCondition.eq}
\end{equation} 

Resolution of equation~\ref{InstabilityCondition.eq} leads to a prediction of the strain at which bucking should appear as a function of sample length $L_y$ for a given mode $n$ (see ~\ref{EpsilonBucklingvsLy.fig}).

In the next section, we present a comparison between our  MD simulations and this linear stability analysis.

\section{ Influence of the sample size}
\label{sec:SampSize}

\begin{figure*}
\begin{center}
\includegraphics[width=0.48\textwidth]{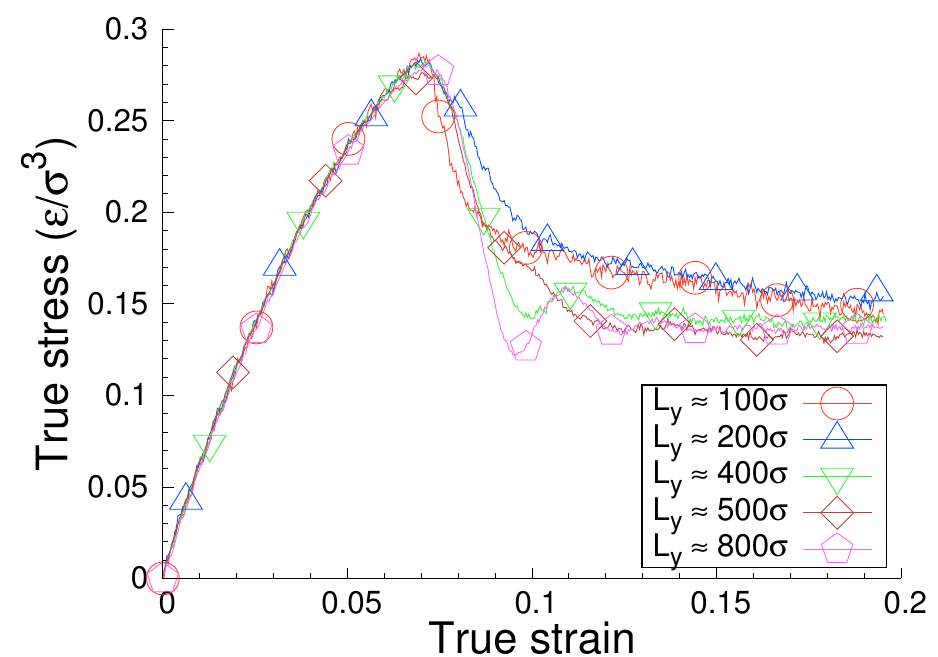}
\includegraphics[width=0.48\textwidth]{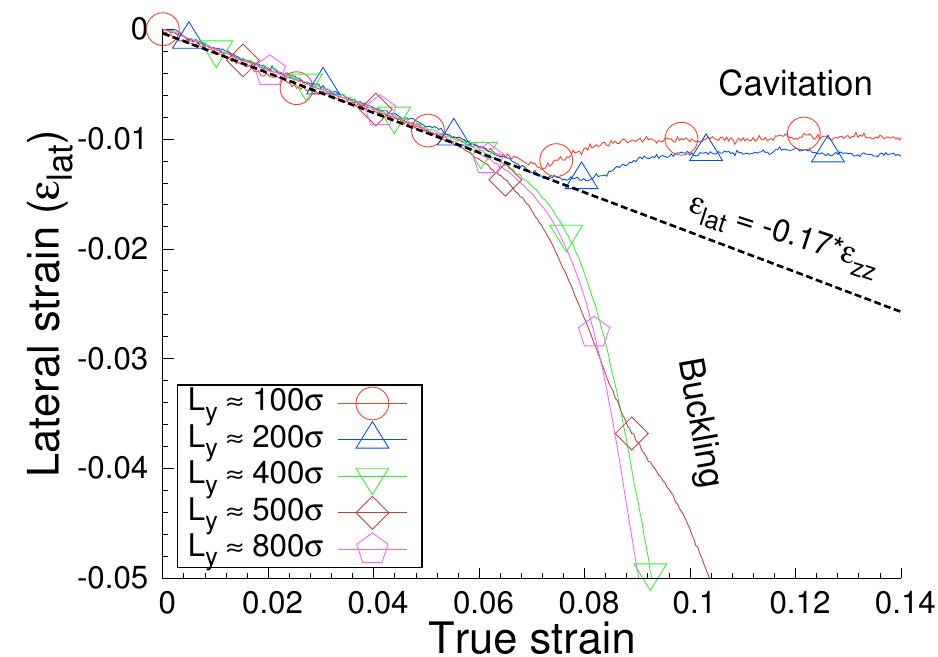}
\end{center}
\caption{ Left panel: Comparison between the mechanical responses of several samples with different sizes, the strain rate is $\dot{\epsilon_{zz}} = 7.3\times 10^{-5}$ for all samples. The corresponding lateral strain curves are shown in the right-panel.  All curves have the same yield point; however, the largest three samples exhibits buckling while only cavitation is present in the  two smaller samples. (see the snapshots in figure \ref{fig:Snapshots}). When buckling is observed, the buckling strain is roughly independent of sample size,  which is in contrast with the prediction from  elastic theory.}
\label{fig:CompareStressStrain}
\end{figure*}


According to the elastic theory, the instability will take place at smaller  strains for bigger samples,
and always at the largest possible wavelength allowed by the boundary conditions. 
 We have therefore studied several samples with different sizes.
These samples were created  by replicating the \emph{same} elementary cell $i$ times along the $Y$ direction, where $i = 3,~6,~12,~15$ and $24$ times.
Because of the periodic boundary conditions, the buckling wave length must be an integer subdivision of the sample size, $k_n = n \times \frac{2\pi}{L_y}$ . 

Figure  \ref{fig:CompareStressStrain} compares the mechanical response of all tested samples, at the same strain rate $\dot{\epsilon_{zz}} = 7.3\times 10^{-5}$.
In terms of stress-strain relation (figure \ref{fig:CompareStressStrain}.a),  all samples have roughly the same mechanical response up to the yield point. The drawing regimes exhibits important differences  between smaller  and larger samples.
 The stress softening in large samples is  more pronounced  than in the small  ones.
Indeed, both cavitation and buckling can limit the stress growth of the elastic regime, leading then to a stress drop.
 For long samples, the two mechanisms  participate in the stress softening, thus the drop of stress will be increased  compared to the short sample case where only  cavitation is present.

The right panel in figure \ref{fig:CompareStressStrain}  shows the lateral strain in the different samples.  All curves  fit very well the same Poisson ratio in the first linear part (dashed line). After the yield strain, strong deviations can be observed: the lateral strain decreases for long samples ($L_y\geq 393.6\sigma$) while it increases for the shorter ones ($L_y\leq 196.8\sigma$).
 The decrease in  lateral strain is related  to  buckling instability, as described previously.
 For short samples, the increase of lateral strain after the yield is correlated with the nucleation of cavities in the rubbery phase. 
 The buckling in such samples is completely absent, as shown in the snapshots of  figure  \ref{fig:Snapshots}.

Examining the behavior of the different samples, one concludes that the minimal length for observing buckling before cavitation is 
 between $196.8 \le L^*_y \leq 393.6$. 
 For samples larger than $393.6\sigma$, the onset of  buckling occurs always  at the same strain ($\epsilon_{buck}=0.06$), in contradiction with the expectation from the elastic description of the previous section.
  A tentative explanation of this behavior will be given below, when we study the influence of the strain rate. 
  Another surprising observation, illustrated in  figure \ref{fig:Snapshots}, is that the wavelength of the instability does not appear to increase with the size of the system, again in contradiction with the expectation  from elastic theory.

\section{Mechanical behavior at lower strain rate}
\label{sec:LSR}
The observations from the section \ref{sec:SampSize} show a difference between the predicted buckling strain and the measures made by simulation.
In order to interpret this difference, we  inspect  in this section the influence of the strain rate on the mechanical response. The same samples described  in the previous section will now be submitted to a similar tensile test, at a strain rate that is smaller by a factor of approximately 5.

\subsection{Stress-Strain response}
\label{ssec:StressStrainLowSR}

\begin{figure}
\begin{center}
\includegraphics[width=0.45\textwidth]{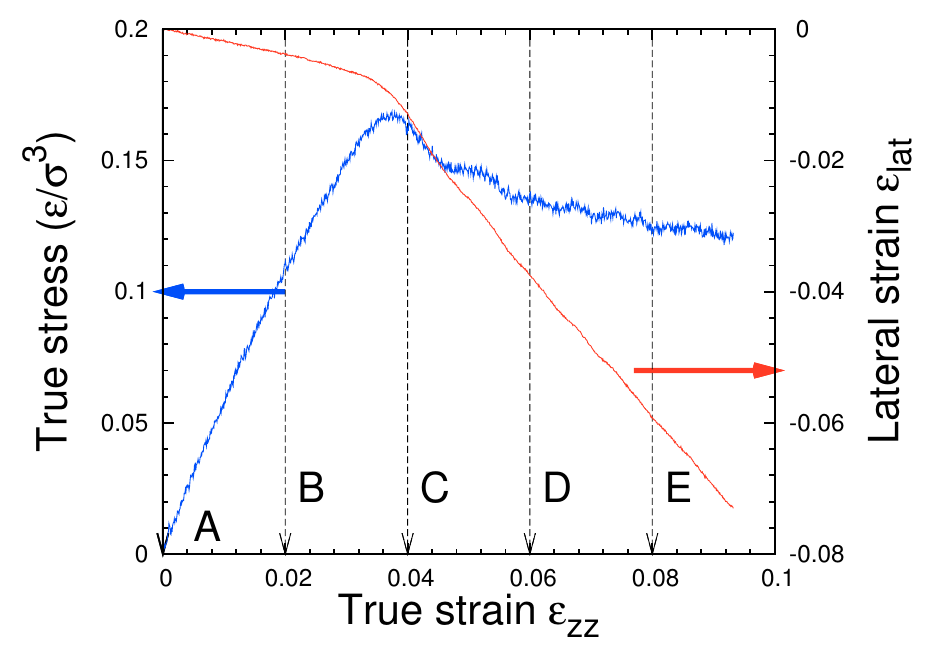}
\includegraphics[width=0.5\textwidth]{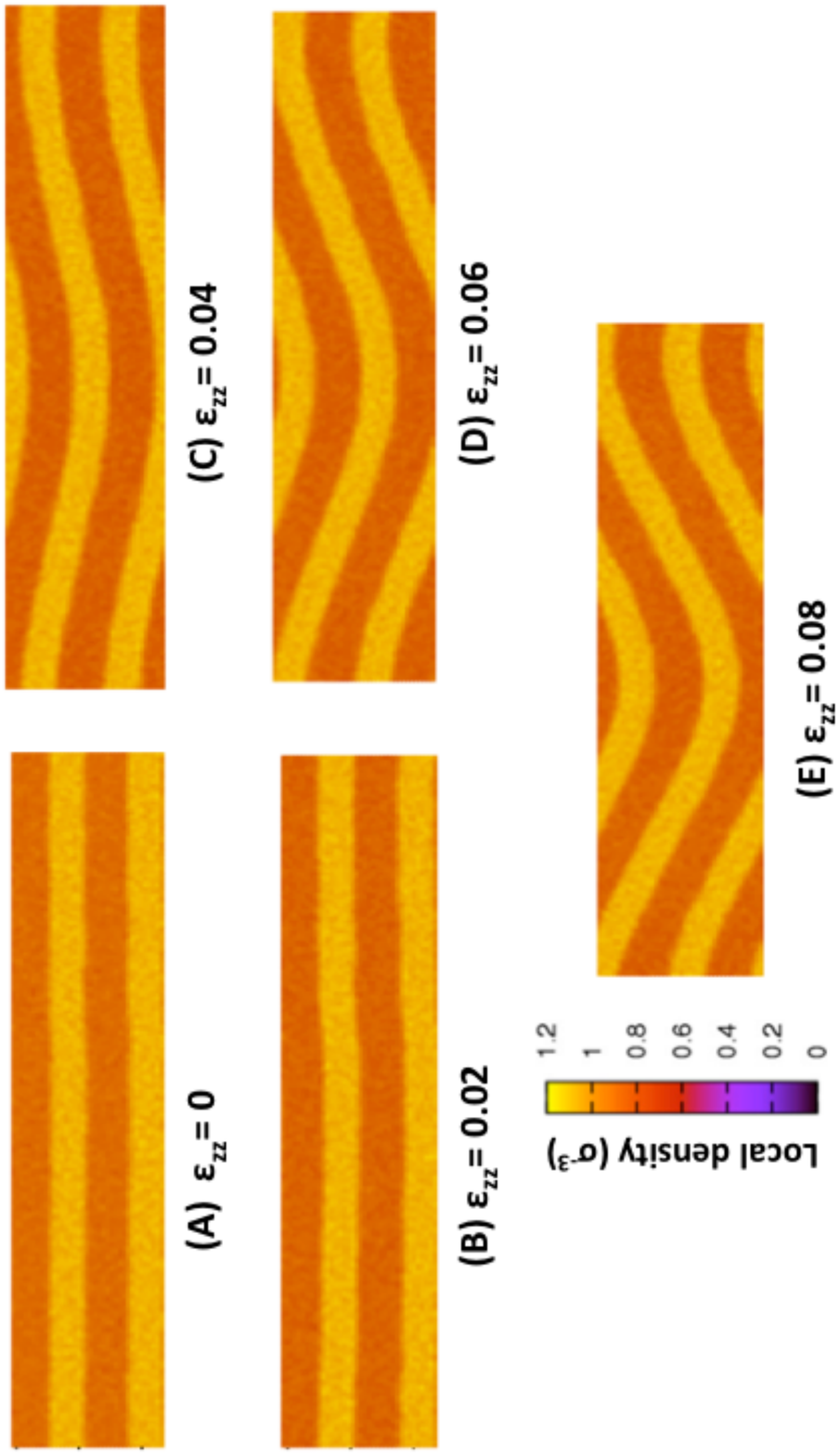}
\end{center}
\caption{Low strain rate tensile test: the upper panel shows the stress strain curve for the same sample illustrated in  figure \ref {fig:BehaviorCurveH}, together with the evolution of  s the lateral strain. In this case, the yield corresponds  to the onset of the buckling.  The arrows indicate the strain levels at which the snapshots presented in the lower panel are taken. Cavities are completely absent from the rubbery phase.}
\label{fig:BehaviorCurveL}
\end{figure}
 \ref{fig:BehaviorCurveL} shows the results of a tensile test performed under the same conditions as in section \ref{sec:StressStrain},  except for
 the strain rate which is  5 times smaller,  $\dot{\epsilon}_{yy} = 1.4\times10^{-5}$.
The resulting stress-strain curve, superimposed with the evolution of  lateral strain, are shown in the top panel.
The  linear part of both curves corresponds to the elastic regime (the stress strain curve fits the Young modulus in this regime and the lateral strain curve fits the Poisson ratio). The end of this regime is marked by the yield, followed by a stress softening in stress-strain curve. 
The yield point corresponds  to the onset of buckling, also  indicated  by the change  of the apparent Poisson ratio in the lateral strain curve.
The absence of cavities was checked  by inspecting the local density of the sample at different strain levels. Therefore, the yield and the stress softening in this case is correlated only to the onset and the development of the buckling in the sample.
The last part of the stress-strain curve is the drawing regime, that corresponds to the development of the buckling undulation  in an ``accordion'' like manner. Note that the range of strain studied here is relatively small, so that the strain hardening regime is not attained.

\begin{figure*}

$\begin{array}{cc}
L_y=196.8\sigma & L_y=787.2\sigma \\
\includegraphics[width=0.3\textwidth]{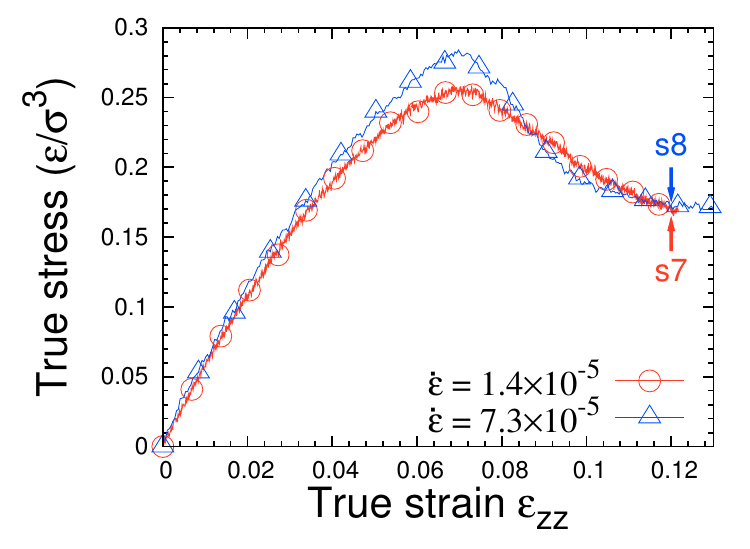} &
 \includegraphics[width=0.3\textwidth]{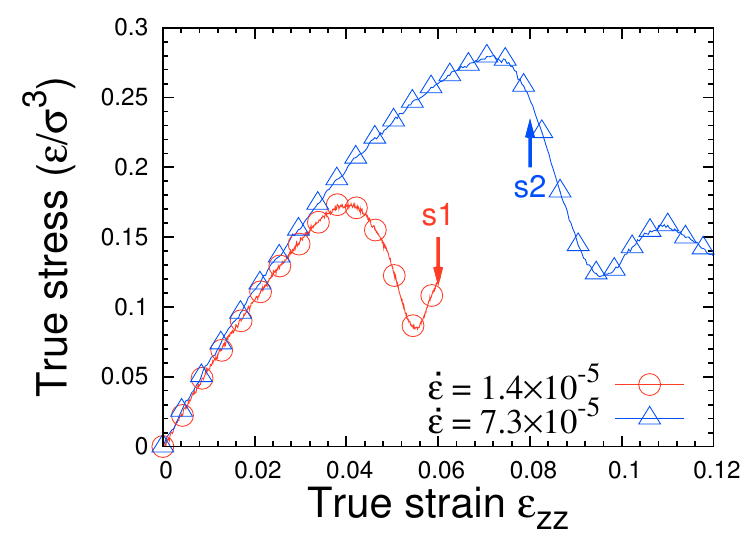} \\
\includegraphics[width=0.3\textwidth]{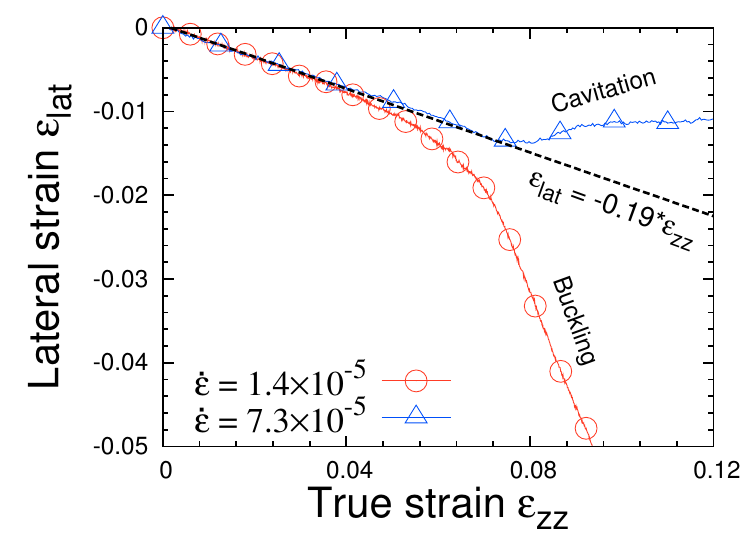} &
\includegraphics[width=0.3\textwidth]{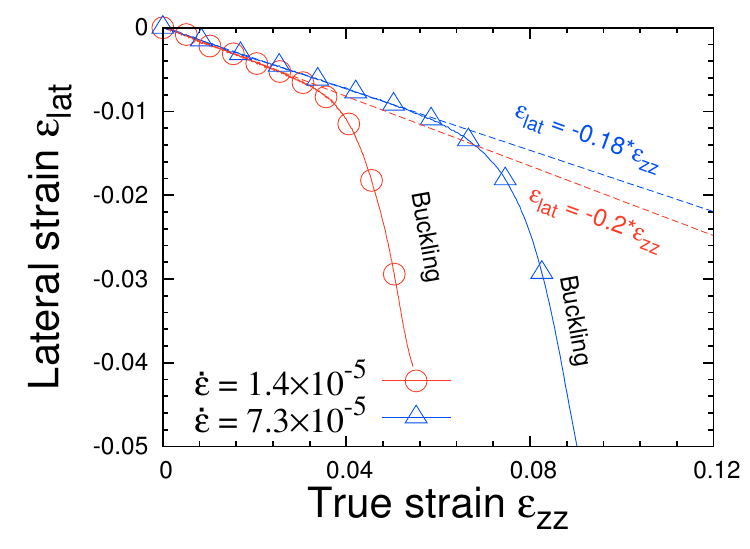} \\
\end{array}$
\caption{ Upper panel: stress-strain curves of the largest and the shortest samples compared at two different strain rates $\dot{\epsilon_{zz}} = 7.3 \times 10^{-5}$ and $\dot{\epsilon_{zz}} = 1.4 \times 10^{-5}$, the lateral strain curves are shown in the lower panel. For the largest sample, decreasing the strain rate will decrease the yield and the buckling strain. This behavior is correlated to the change of the buckling mode as illustrated in  figure \ref{fig:Snapshots}. For the shortest sample, the behavior changes from cavitation to buckling (see figure   \ref{fig:Snapshots}) .  }
\label{fig:CompareBehavior}
\end{figure*}

\subsection{Influence of  sample size and of strain rate}

In figure \ref{fig:CompareBehavior}, we compare the stress-strain and the lateral strain curves of our smaller and larger samples, for the two strain rates under consideration.  Clearly the Young modulus is essentially independent of system size and strain rate. 
In contrast, the yield stress and strain decrease as the strain rate decreases, most markedly  in the larger sample.  Finally, the stress softening is significantly weaker at at low strain rate,  again especially in the large sample.

\begin{figure*}
\includegraphics[width=0.7\textwidth]{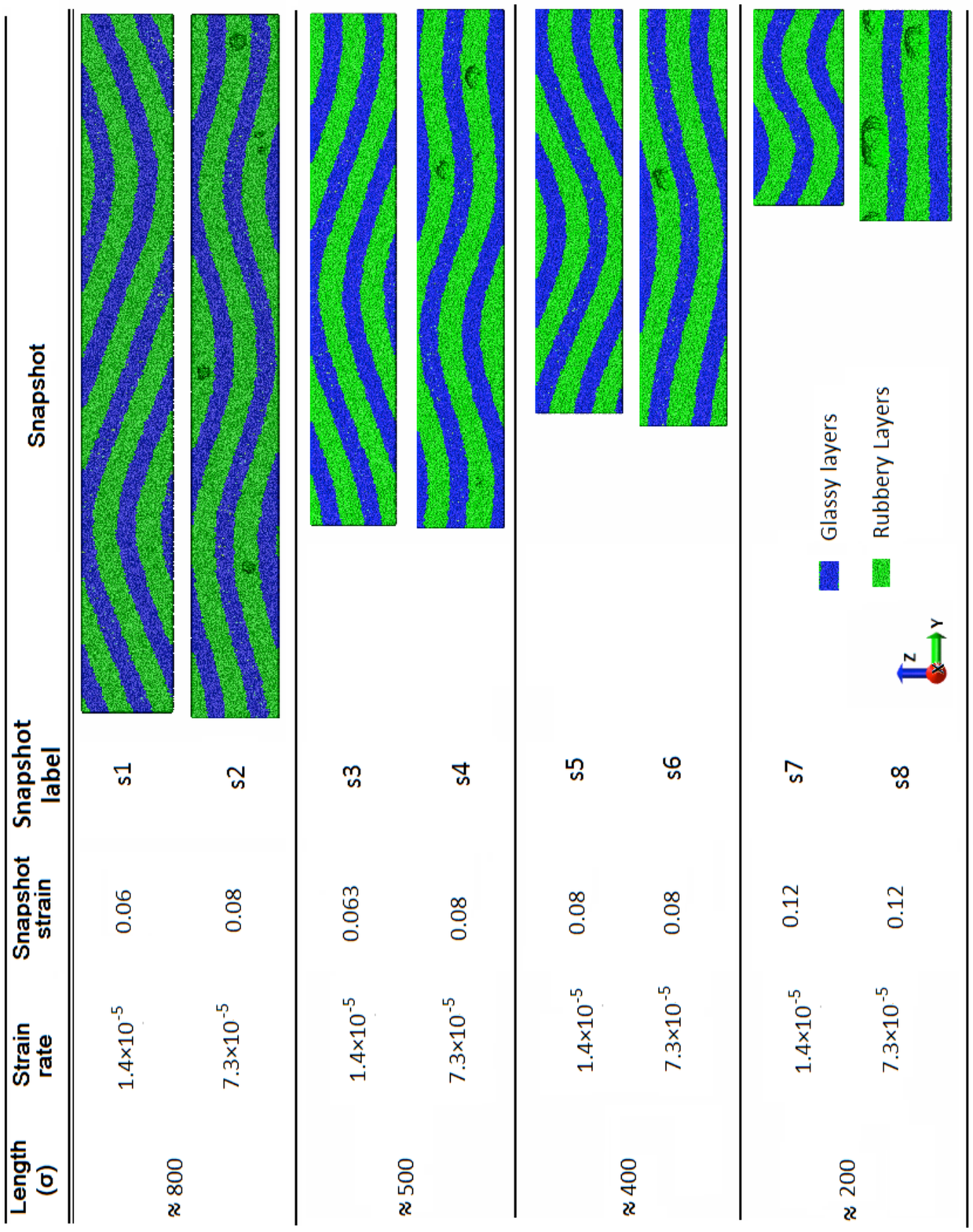}
\caption{ Snapshots show several samples under a uniaxial tensile tests driven by two different strain rates $\dot{\epsilon_{zz}} = 7.3 \times 10^{-5}$ and $\dot{\epsilon_{zz}} = 1.4 \times 10^{-5}$ Several lengths are presented, at low strain rate all samples buckle. The buckling wave length is equal to the sample length. At high strain rate, the buckling wave length seems to be independent from the sample length. The labels correspond to those indicated in figure  \ref{fig:CompareBehavior}}
\label{fig:Snapshots}
\end{figure*}

 In general, the decrease of the yield threshold is strongly correlated with the change of the plastic deformation mode from cavitation to buckling\footnote{The occurrence of buckling for all samples at low strain rate is illustrated in figure \ref{fig:Snapshots}}. Both cavitation and buckling result in a yield behavior, however the yielding associated with buckling is much more progressive and smooth than the one associated with cavitation.
For the smallest sample  ($L_y=196.8\sigma$), the lateral strain curve  also highlights a radical change of the mechanical response form cavitation to buckling at low strain rate.

Figure \ref{fig:Snapshots} compares the configurations after buckling,  at two different strain rates.
The change of the yield mechanism from cavitation to buckling is well illustrated in these snapshots especially for the smallest sample. The second important observation is that the wavelength becomes equal to the sample length in all samples at low strain rate. Finally, there are no cavities present in the rubbery phase of the lower strain rate configurations, compared to systems deformed at high strain rate  for the same strain.
These snapshots confirm that the low energy  buckling mode is adopted by the sample at the lowest strain rate. 

\begin{figure*}
\includegraphics[width=0.7\textwidth]{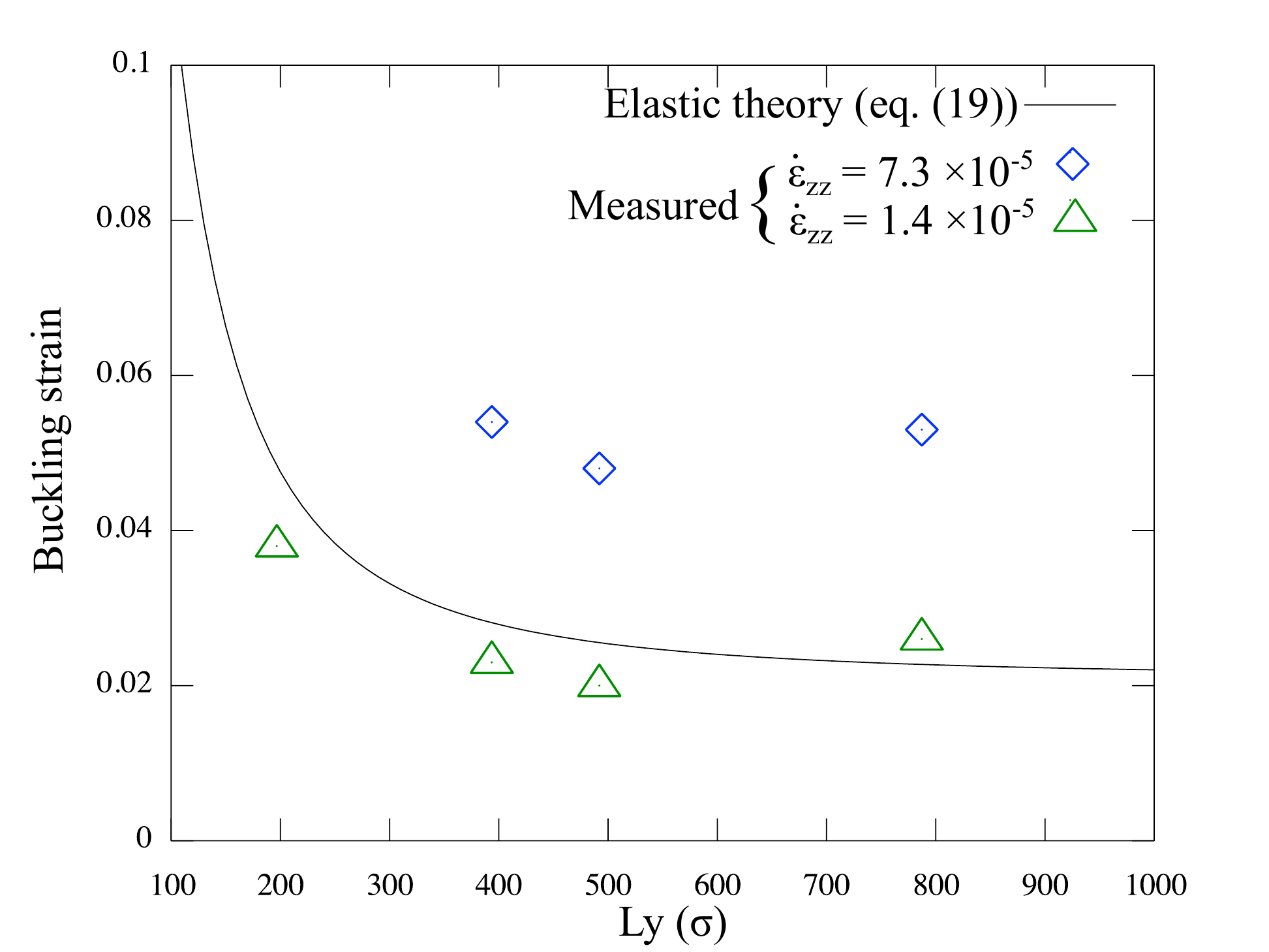}
\caption{Comparison between the buckling strain values predicted from the elastic theory (\ref{InstabilityCondition.eq}) and the one measured from MD simulations for all samples at two strain rates $7.3\times10^{-5}$ and $1.4\times10^{-5}$.}
\label{EpsilonBucklingvsLy.fig}
\end{figure*}
\ref{EpsilonBucklingvsLy.fig} compares the buckling strain values predicted from the elastic theory and the one measured from MD simulation for all samples at two strain rates. The MD buckling strain is defined by the value of the strain at which a deviation from the linear Poisson behavior is detected in the lateral strain. Elastic theory predictions (\ref{InstabilityCondition.eq}) correspond well to MD simulation performed at low strain rate, which is not the case with  high strain rate MD simulations.   

Summarizing these observations, the role of the strain rate seems to be determinant for the mechanical response of the sample.
 Depending on the applied rate of the deformation, the samples switch between the fundamental and the second mode of buckling or between the cavitation and buckling. In the next section, a simple model  will be proposed to account for this dependence of the buckling instability on strain rate.

\subsection{Unloading process}

\begin{figure*}
\begin{center}
\includegraphics[width=0.48\textwidth]{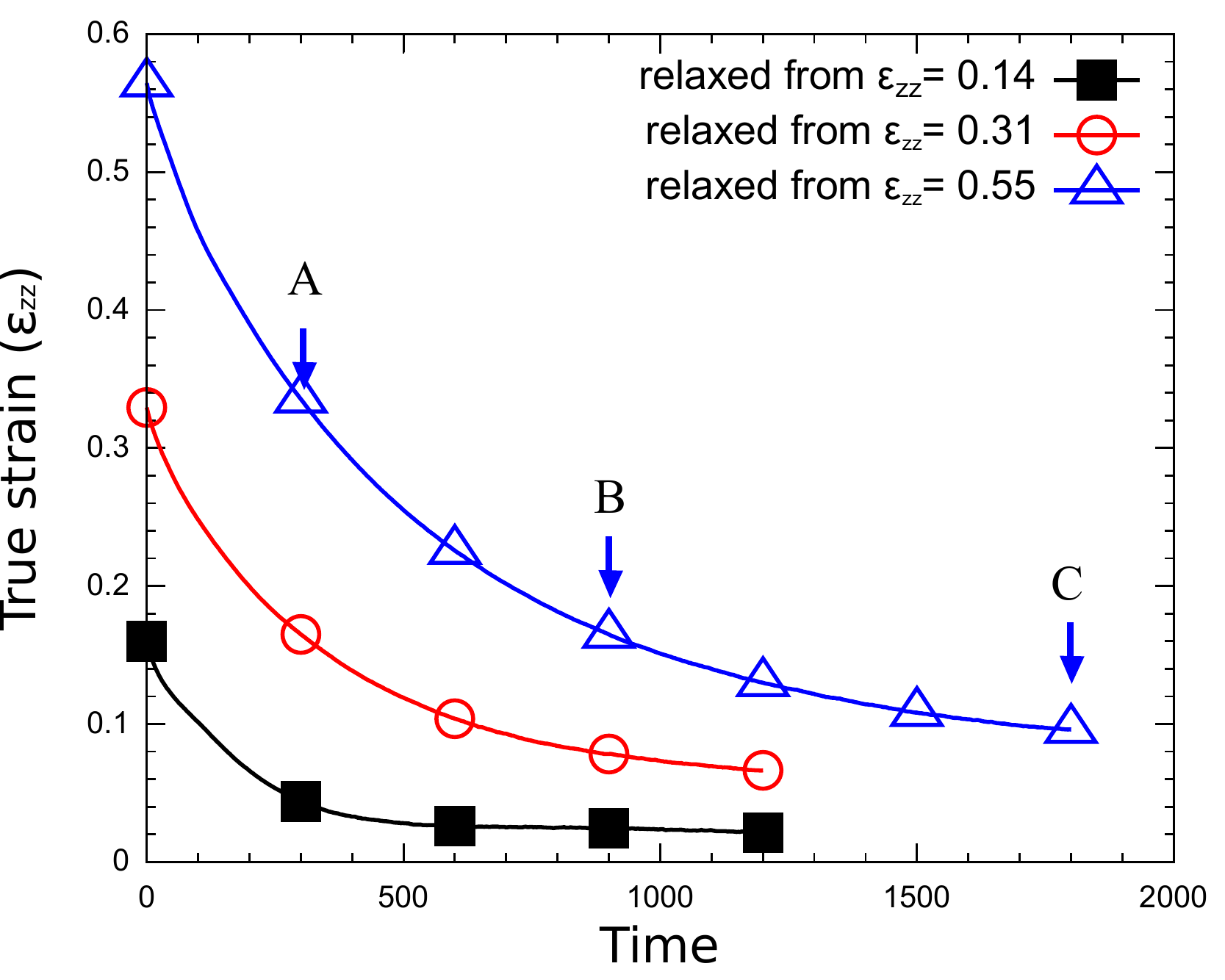}
\includegraphics[width=0.48\textwidth]{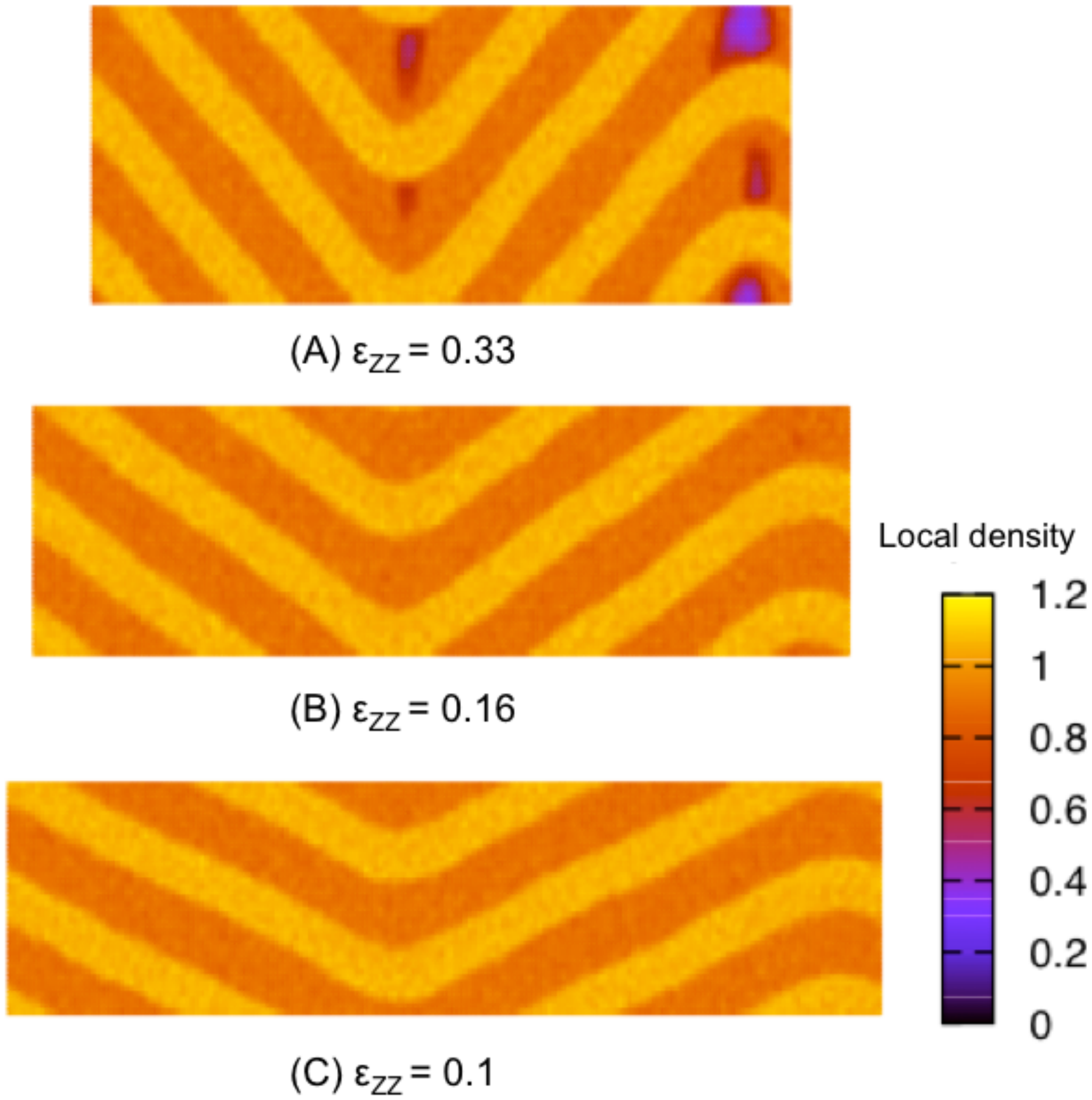}
\end{center}
\caption{Left: relaxation of the strain in a system unloaded instantaneously after being deformed to the strains indicated in the figure. Right : snapshots of the relaxing configuration (density field) at the three states indicated by A,B, C in the left panel.}
\label{fig:relax}
\end{figure*}

The irreversible aspect of the deformation was studied by instantaneously unloading the sample after 
deforming to different final strains, and monitoring the subsequent relaxation of the strain. The results of these relaxation simulations are shown in figure  \ref{fig:relax}. It is seen that the process is clearly irreversible only for the largest deformation. For such deformations, significant cavitation has taken place at the hinges of the chevron pattern. The resulting density pattern under zero load displays a chevron structure with quite large angles, as observed in experiments. However, the residual, irreversible deformation is quite small, about 10\% for a total deformation of 50\%. This is presumably the consequence of a model in which the 'hard' phase remains in fact relatively soft (i.e. the contrast in moduli is smaller than in experiments) and irreversible damage implying chain breaking is excluded. As a consequence  the main source of  irreversible damage in the glassy  phase is the cavitation process. 

\section{Competition between buckling modes}
\subsection {Modeling}

\begin{figure*}
\begin{center}
\includegraphics[width=0.48\textwidth]{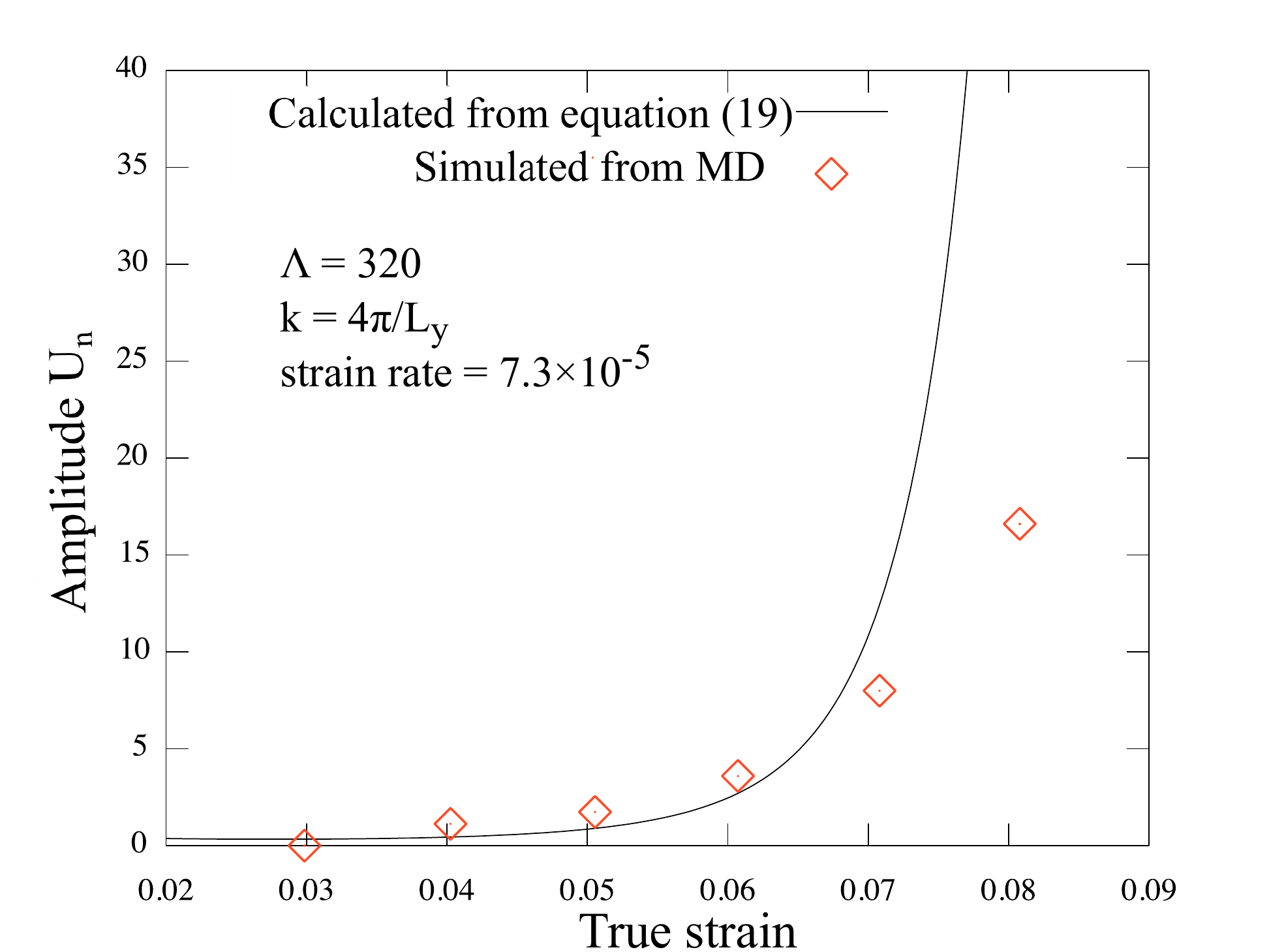}
\includegraphics[width=0.48\textwidth]{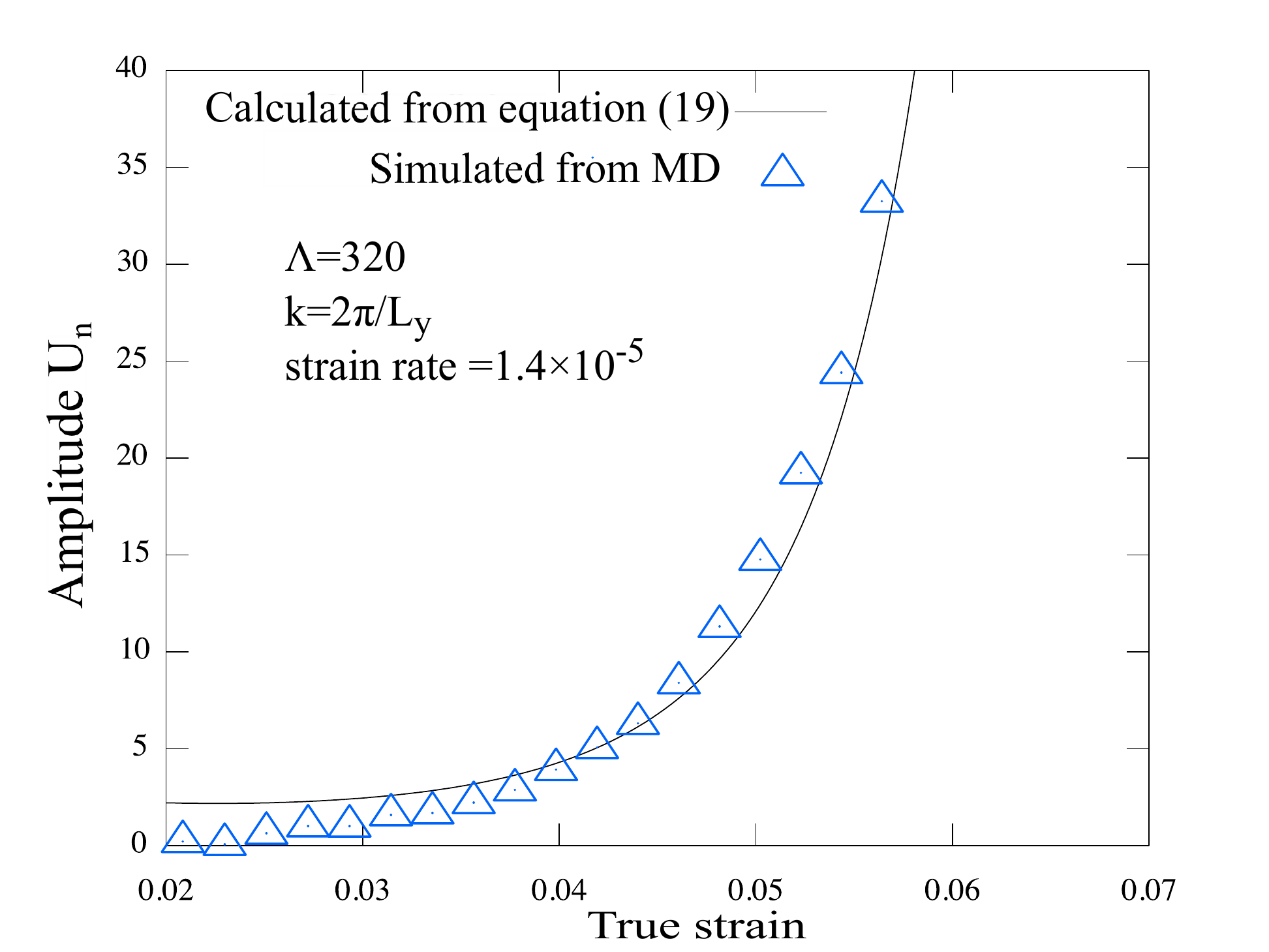}
\end{center}
\caption{Comparison between the measured values of buckling amplitude (disconnected symbols) and the value calculated from equation 9 (full lines) for the largest sample $L_y = 787.2\sigma$. The left panel corresponds to fundamental mode $k = \frac{2\pi}{L_y}$ observed at low strain rate, whereas the right panel corresponds to the second mode $k = \frac{4\pi}{L_y}$ at high strain rate.  The value of $\Lambda$  taken here corresponds to the best fit of the measured amplitude. }
\label{fig:FitLambda}
\end{figure*}

The results from the previous sections show a unexpected change of wave vector with strain rate.    
To understand  this observation, we propose to describe the growth of the buckling amplitude $U_n(t)$ for  a wave-vector $k_n=2n\pi/L_y$
using a simple linear relaxation equation of the form:
\begin{equation}
\label{kinetic.eq}
\frac{dU_n}{dt} =  -\Lambda .F(\epsilon_{zz},k_n) . U_n
\end{equation}
where $F(\epsilon,k_n) . U_n$ is the driving force. 
$\Lambda$ is 
a phenomenological coefficient which will be assumed to be independent from wave-vector and the strain rate. 
The  solution of this equation can be written as:
\begin{equation}
\label{ana_sol_kinetic.eq}
U_n(t) =  U_n(0).\exp \left(-\Lambda .\int_0^t {F(\epsilon(s),k).ds} \right)
\end{equation}
Note that the strain  $\epsilon_{zz}$ is a time dependent variable $\epsilon_{zz}(t) = \dot{\epsilon_{zz}} \times t$.
One also remarks that  $U_n(t)$   in equation \ref{ana_sol_kinetic.eq}  is not monotonous. The function 
passes trough a minimum at short times (small strains) , as the phenomenological equation \ref{kinetic.eq} is purely relaxational. For large enough strains, the growth rate becomes positive, and $U_n(t)$ grows exponentially.
The growth starts when $\epsilon_{zz}$ corresponds to the buckling strain ($\epsilon_{zz}= \epsilon^*_{zz}(k)$) for which  $F(\epsilon_{zz}(t),k) = 0$. The decreasing part of the curve prior to buckling is irrelevant, as thermal fluctuations are ignored in equation  \ref{kinetic.eq} .

 
Equation \ref{kinetic.eq} has been solved numerically using a fourth order Runge-Kutta method  starting from an arbitrary small value of $U_n(0)$.
The value of $\Lambda$ and $U_n(0)$ were determined by the solution that ensures the best fit of the buckling amplitude measured during
 the deformation of the largest sample $S_{24}$  at low and high strain rate (see  figure \ref{fig:FitLambda}).
Note that the largest sample was chosen for the fit because the two buckling modes $2\pi/L_y$ and $4\pi/L_y$ can be clearly distinguished at two different strain rates.
We find that  the best fit is obtained for $\Lambda=320$ and the initial value of $U_n(0)=0.03$.
This initial value of the amplitude can be considered as corresponding to the level of heterogeneity of the  initial sample, which serves to initiate the buckling.

\subsection{Results and discussion}

\begin{figure}
\begin{center}
\includegraphics[width=0.48\textwidth]{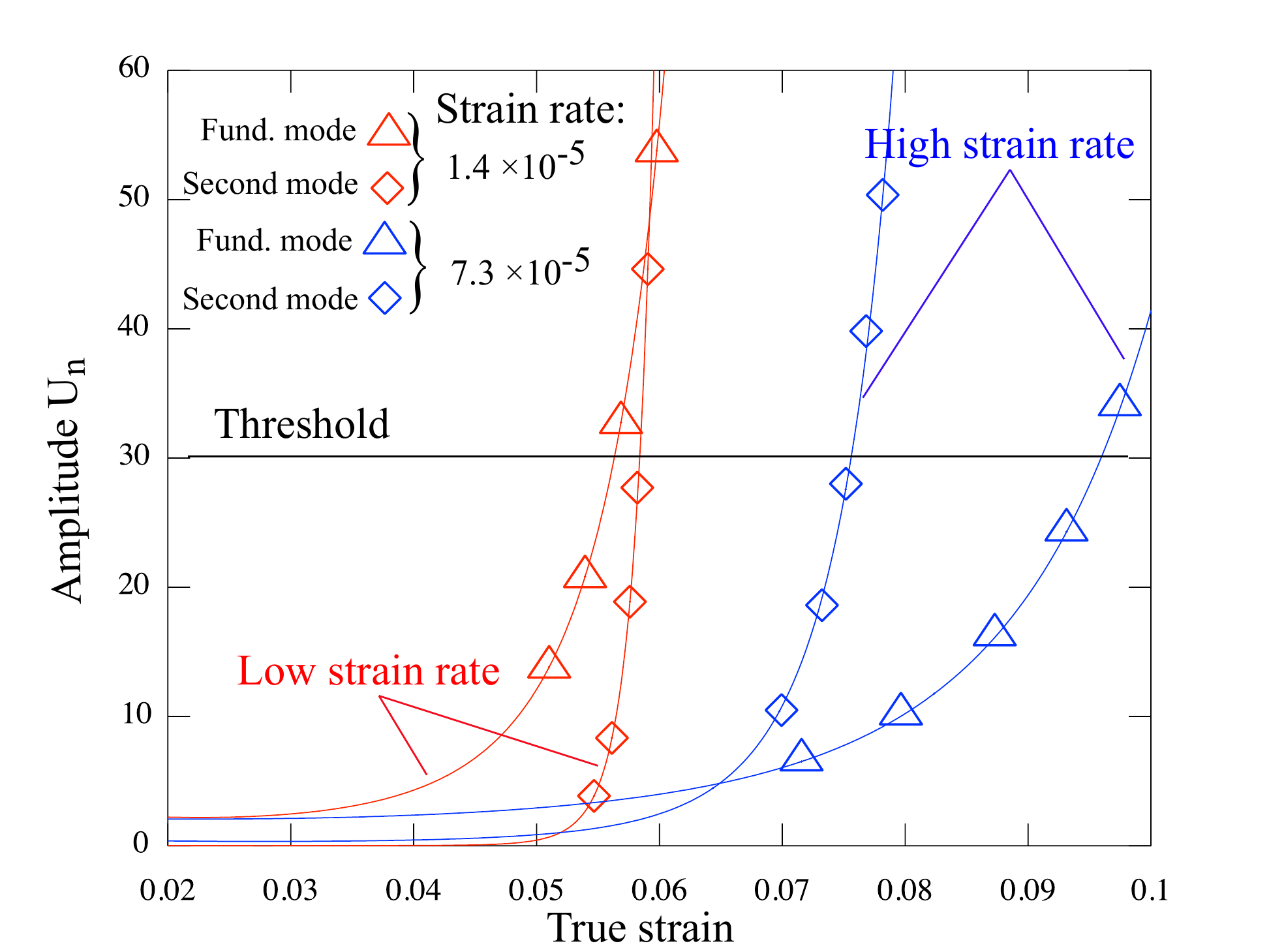}
\end{center}
\caption{ Comparison between the growth of the  fundamental mode and of  the second  mode for the largest  sample at two different strain rates. The higher mode develops faster than the lower mode even if the instability of the latter is triggered at first. The threshold is chosen so that it intersects first  the fundamental mode at low strain rate, and the second mode at high strain rate. }
\label{fig:Threshold}
\end{figure}

Having fixed $\Lambda$ and $U_n(0)$, equation \ref{kinetic.eq} has been solved for different values of wave vectors $2\pi/L_y$ and $4\pi/L_y$ at two different strain rates $\dot{\epsilon_{zz}} = 7.3\times 10^{-5}$ and $\dot{\epsilon_{zz}} = 1.4\times 10^{-5}$. 
The results  were plotted and compared in figure  \ref{fig:Threshold}.
This figure shows that (i) the buckling strain increases as the buckling wave vector increases, as expected; (ii) at the same strain rate, the second buckling mode is  faster to develop, compared to the fundamental mode.
Moreover the second mode growth intercepts the fundamental buckling growth at a certain strain (called below switching strain,  $\epsilon_{sw}$) for both strain rates. At this strain the second buckling mode can overtake  the fundamental mode if the latter has  not  fully developed.
For the low strain rate case, the fundamental buckling mode starts growing at a low strain, and is well developed when the strain reaches  $\epsilon_{sw}$, so that the system can not switch to higher buckling mode, and the fundamental mode is selected. 
In contrast, at high strain rate the first mode has grown to a small amplitude when the sample reaches $\epsilon_{sw}$ , 
so that the second mode can easily overwhelm the fundamental mode and the sample adopts a higher mode for buckling.

From the previous analysis one can define a critical buckling amplitude: as the amplitude threshold after which the selection of the  buckling mode is prohibited.
This mean that the buckling mode that reaches this threshold at the first is the one that is adopted by the sample to achieve the buckling. This mode is called hereafter the ``winner mode''. Quantitatively speaking the ``winner mode'' of buckling is defined when the strain elapsed to reach the critical amplitude is minimum.

 The critical amplitude threshold can be determined approximatively from figure  \ref{fig:Threshold}. This threshold graphically located in the middle of the interval delimited by the two switching points\footnote{This term designates the switching from the fundamental to the second mode only} for each strain rate curves. For the rest of analysis the amplitude threshold was  taken $U^*= 30\sigma$, at this value the threshold crosses the fundamental mode growth at first at low deformation rate while it crosses the second mode at first at high deformation rate.
 
 Obviously, the procedure that consists in   fixing the amplitude threshold at which a mode will become predominant is empirical. The actual mechanism for mode selection presumably involves nonlinear interactions between modes, which are not accounted for in the present description. 


\begin{figure}
\begin{center}
\includegraphics[width=0.48\textwidth]{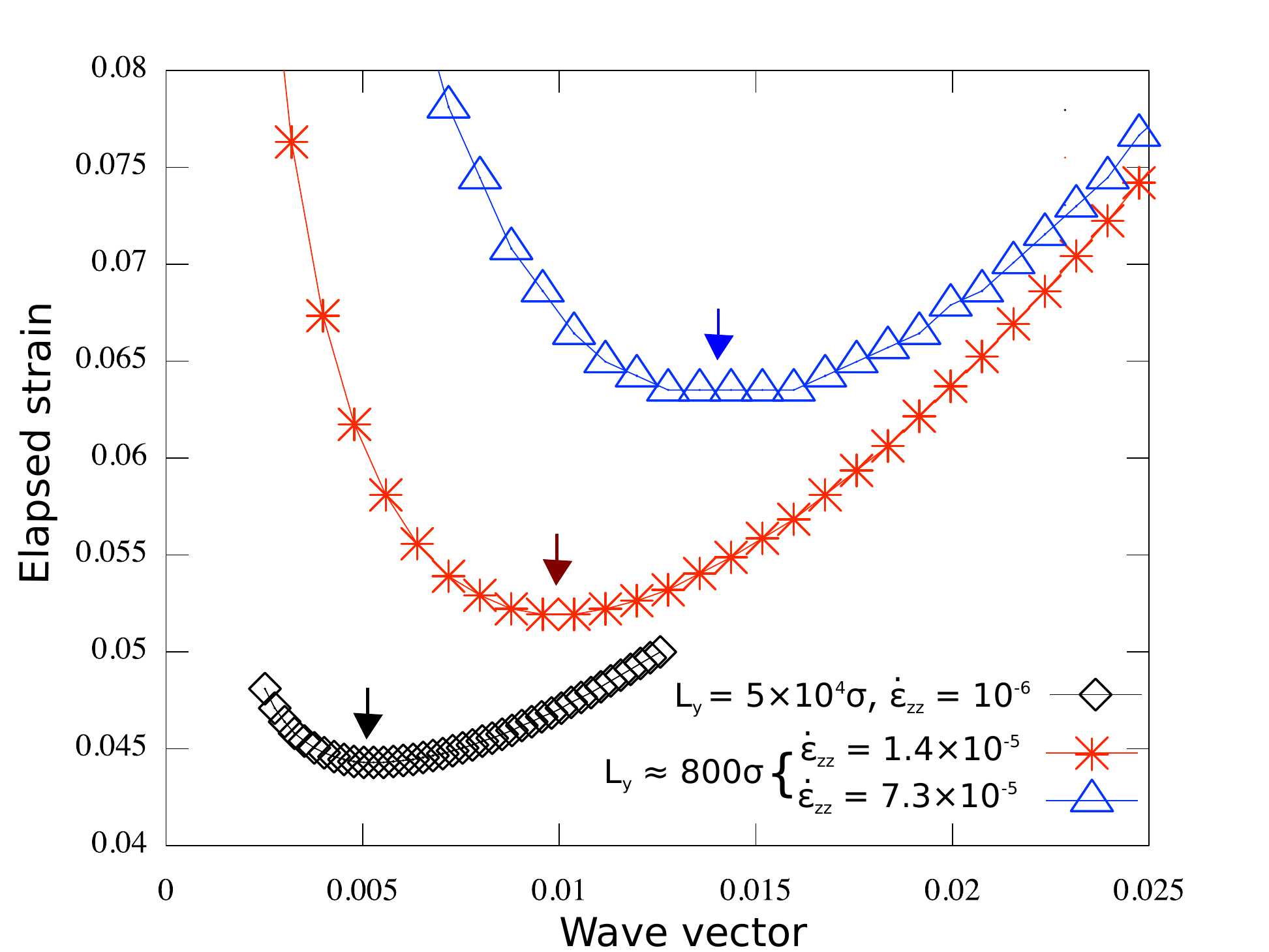}
\end{center}
\caption{ True strain elapsed from the beginning of the deformation to reach the threshold for different buckling wave vectors, according to the kinetic model : the minimum corresponds to the ``winner mode'', that will be adopted by the sample. Several different system sizes and strain rates varying over two orders of magnitude are considered.}
\label{fig:MinimalK}
\end{figure}

 \subsection{Generalization}
 In order to generalize the previous analysis: let one consider a large sample that is deformed at an imposed  strain rate $\dot{\epsilon_{zz}}$.
The choice of a large sample size leads to very close values of wave vectors $2\pi/L_y \simeq 4\pi/L_y \simeq 6\pi/L_y$... .
To identify the mode that is likely to be selected, one has to compute  the growth of the amplitude for each  mode, which was done using the same   parameter $\Lambda$  as above. 
The strain needed  for the amplitude of a mode  to reach the  threshold $U^*$ can then be  computed as a function of $k$.
The result is plotted in figure  \ref{fig:MinimalK}.
The curve passes trough a minimum that separates two regimes:
the decreasing portion of the curve that corresponds to the fact that the development of buckling becomes faster as the buckling wave vector increases.
The second regime, after the minimum,  corresponds to the increase of the buckling strain observed when the buckling wave vector increases.
The value of the minimum gives the wavevector that will be selected at the strain rate under consideration. Besides the approximation of using a wavevector independent threshold, 
this prediction also ignores   the nucleation of cavities observed at extremely high strain rate, and neglects any variation of $\Lambda$ with respect to the deformation rate.


\section{Conclusion}
\label{sec:Conclusion}

In this paper, the mechanical response of triblock copolymer models in the lamellar phase has been investigated by using a coarse grained molecular dynamic simulation.
 Our MD samples were built by radical like polymerization method, and alternate glassy and rubbery lamellae.
Uniaxial tensile tests were performed in the  direction normal  to the layers. 
The resulting constitutive laws are compared to the change of sample morphology and microstructure.
 
 Depending on the applied deformation rate,  the  samples exhibit a variety of microscopic deformation mechanisms. 
   At relatively high strain rate, one observes (except for the shortest samples) a buckling of the lamellae into a wavelength that does not depend on the sample size. 
   The buckling is accompanied  by the nucleation of cavities in the rubbery phase. Both events were found to contribute to the drop stress at yield.
  At low strain rate, all samples (including the shortest one) exhibit buckling. The yield becomes correlated with the onset of the lamellae buckling, and the cavitation is delayed. 
  The undulation wavelength is almost equal to the sample length in this case.
  The buckling strain of each sample was calculated using the elastic theory approach. The results were compared to the  values observed  in the MD simulations.
   A strong deviation was found at high strain rate; however, at low strain rate, the results are consistent. 

This behavior was rationalized by using a simplified  model of mode growth based on  a viscous dynamics and on an elastic driving force for the mode amplitudes.
This model shows that the higher mode of buckling is  faster to develop compared to the fundamental mode, although the latter is the first to become unstable. When the strain rate increases, several modes come into competition.
 The shortest wavelength that corresponds to a larger driving force can take over and dominate the instability pattern.
  In this case the strain for observing buckling can be markedly larger than predicted by elastic theory.

 We finally turn to a short discussion of the relevance of our results to experimental situations. Our simulations use a coarse grained model, which is not specific to any material, and is defined in terms of typical energy, mass and size.  Using an energy scale of $1000 K\times k_B$ and a length scale of 0.5 nm, which are typical in the coarse-grained descriptions of standard polymers, the corresponding stress unit is of order 100 MPa, and the Young modulus of the glassy polymer is of the order of 1-10 GPa.
The timescale that results from these choices of units, if parameters appropriate for typical polymers are used, lies in the picosecond time range. Therefore, the strain rates achieved in simulations are of the order of $10^7 s^{-1}$ in real units, extremely high compared to typical experimental rates. As is often the case in simulation studies involving glassy materials, the behavior observed in simulation studies must be understood as being qualitatively, rather than quantitatively, representative of the experimental reality. However,  there are several arguments that are indicative of the relevance of the mechanism observed and modeled in this paper to experimental situations. First of all, it is known from experiments that plasticity is usually observed for strains of at least 5\%. For such values of the strain, the wavelength that are energetically favorable within the elastic theory are quite small, as can be seen from figure \ref{fig:ElasticTheory}, and typically in the 100 nm range. The issue is then to understand, why such wavelengths are selected within the experimentally slow deformation process, as opposed to larger wavelengths that should be selected in a truly quasistatic approach. Our molecular simulations and the associated kinetic model are a good indicator here.  Molecular simulations indicate that the critical strain predicted by the static elastic theory is observed for the  lowest strain rates that can be achieved in simulation, which are still  very high compared to experiments. These results are rationalized on the basis of a kinetic model, and the extrapolation to lower strain rates using the kinetic model indicates that the preferred wavelength will be weakly sensitive  to strain rate over a broad range of values, with a selected wavelength that increases by only a factor of 2 for a change of more than one order of magnitude in the strain rate (see figure  \ref{fig:MinimalK}).  Unfortunately the limitations of the model, which considers a strain rate independent kinetic coefficient, does not allow us to extrapolate reliably to experimental strain rates, Another qualitative prediction that can be made on the basis of our analysis concerns the temperature dependence of the chevron pattern; as the kinetic coefficient $\Lambda$ is expected to increase with temperature ($1/\Lambda$ can be associated with a viscosity), the threshold for instability will be reached earlier  for 
larger wavelength. Therefore the wavelength is expected to increase with increasing temperature as it does with decreasing strain rate, in line with general time-temperature considerations.

 \begin{acknowledgement}
Computational support was provided by the Federation Lyonnaise de Calcul Haute
Performance and GENCI/CINES . Financial support from ANR project Nanomeca is
acknowledged. JLB is supported by Institut Universitaire de France. Part of the simulations were carried out using the
LAMMPS molecular dynamics software (http://lammps.sandia.gov), The snapshots were visualized using VMD software\cite{HUMP96} (http://www.ks.uiuc.edu/Research/vmd/) .
\end{acknowledgement}


\bibliography{buckling_detailed_version7.bib}
\begin{tocentry}

\includegraphics[width=0.8\textwidth]{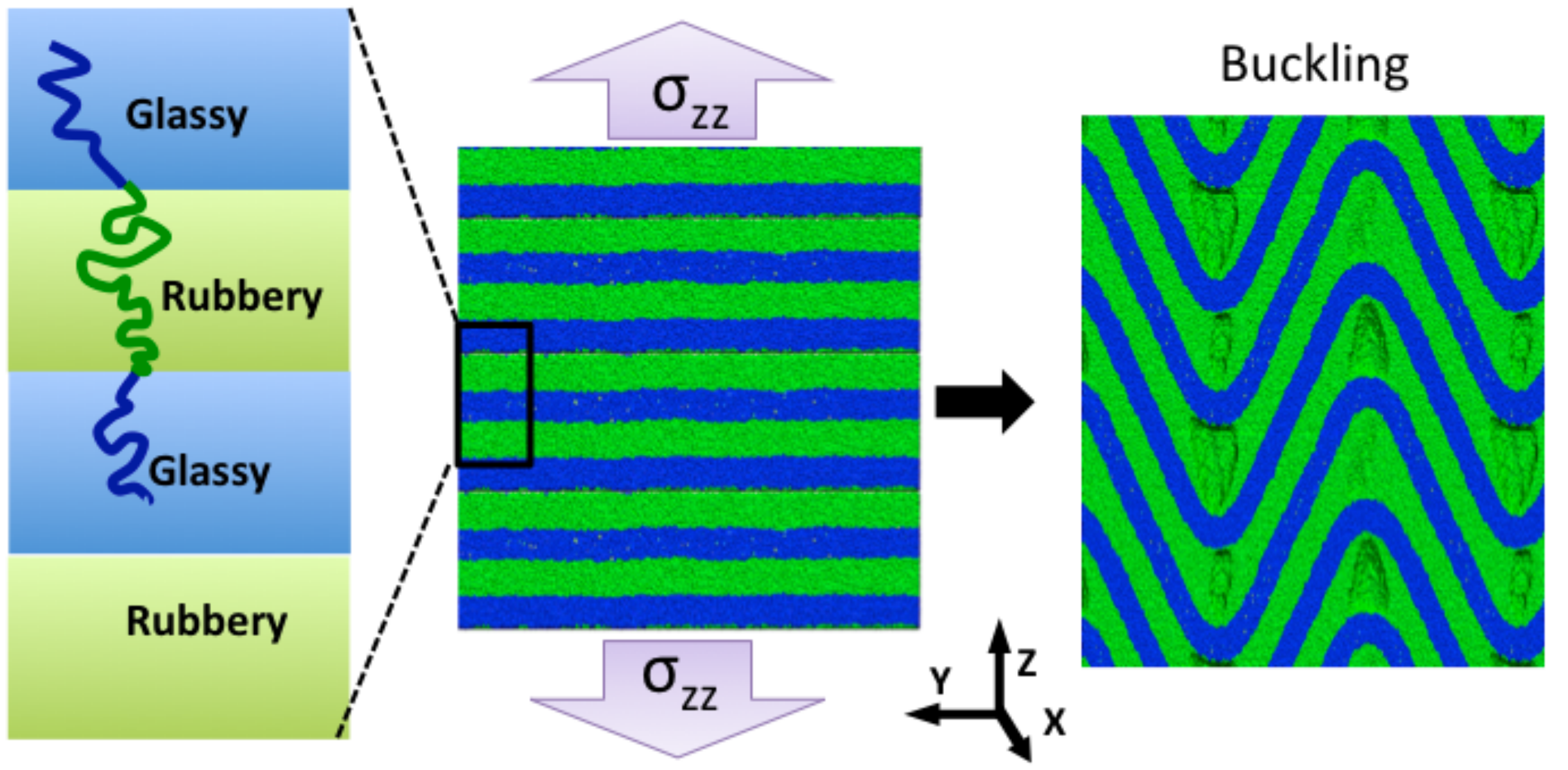}

\end{tocentry}

\end{document}